%% file: main.tex
\pdfoutput=1
\documentclass[letterpaper,twocolumn,10pt]{article}
\usepackage{usenix}

\usepackage{amsthm}
\usepackage{graphicx}
\usepackage{array}
\usepackage{booktabs}
\usepackage{multirow}
\usepackage{xspace}
\usepackage{listings}
\usepackage{pifont}
\usepackage{tcolorbox}

\theoremstyle{definition}
\newtheorem{definition}{Definition}[section]

\newlength{\figwidth}
\newcommand{\circled}[1]{\textcircled{\footnotesize#1}}
\newcommand{\code}[1]{\texttt{#1}}

\newtcolorbox{findingbox}[1][]{
  colback=blue!3,
  colframe=blue!50!black,
  boxrule=0.6pt,
  arc=3pt,
  left=6pt, right=6pt, top=4pt, bottom=4pt,
}

\newcommand{\rqbox}[2]{%
  \par\noindent\textbf{#1:} \textit{#2}\par\smallskip
}

\newcommand{\sysname}{ClawHub\xspace}
\newcommand{\toolname}{\textsc{SkillTriage}\xspace}
\newcommand{\numskills}{48,682\xspace}
\newcommand{\numauthors}{17,130\xspace}
\newcommand{\toolleakratepct}{5.16\%\xspace}
\newcommand{\eg}{\textit{e.g.}\xspace}
\newcommand{\etal}{\textit{et al.}\xspace}
\newcommand{\orcidname}[2]{\href{https://orcid.org/#1}{\color{black}#2}}

\hypersetup{
  pdftitle={Carried-Over and Agent-Native Threats in the Supply Chain of LLM Agent Skills},
  pdfauthor={Kaixuan Li, Wenbo Guo, Sen Chen, Yang Liu}
}

\begin{document}

\date{}

\title{\Large \bf Carried-Over and Agent-Native Threats in the\\ Supply Chain of LLM Agent Skills}

\author{
{\rm \orcidname{0000-0002-3517-353X}{Kaixuan Li}$^{1}$, \orcidname{0000-0001-6655-8179}{Wenbo Guo}$^{1}$, \orcidname{0000-0001-9477-4100}{Sen Chen}$^{2}$, \orcidname{0000-0001-7300-9215}{Yang Liu}$^{1}$}\\
$^{1}$Nanyang Technological University \qquad $^{2}$Nankai University\\
{\normalsize\rm\texttt{kaixuan.li@ntu.edu.sg, wenbo002@e.ntu.edu.sg, senchen@nankai.edu.cn, yangliu@ntu.edu.sg}}
}

\maketitle

\input{sections/abstract}

\input{sections/introduction}
\input{sections/background}
\input{sections/studydesign}
\input{sections/study}
\input{sections/discussion}
\input{sections/relatedwork}
\input{sections/conclusion}

\bibliographystyle{plainurl}
\bibliography{ref}

\end{document}

%% file: sections/abstract.tex
\begin{abstract}
\label{sec:abstract}
Agent skills are reusable packages that extend a large language model (LLM) agent with scripts, third-party dependencies, and natural-language instructions.
An agent builds a skill's arguments from user requests and untrusted external content, so a flaw left by a benign author can escalate a prompt injection to code execution.
Existing studies target malicious skills or a single attack vector, and none tests whether an agent can reach such flaws.

We present the first registry-scale empirical study of the supply chain attack surface arising from unintentional vulnerabilities in agent skills, over all \numskills skills in \sysname.
The study covers two carried-over risks, dependencies and first-party code, and two agent-native ones, install instructions and credentials printed to stdout.
We screen all skills by static analysis, manually review all 3{,}428 cases flagged by the four main detectors, and test reachability by running sampled flagged skills in a sandbox and with a live LLM agent under indirect prompt injection.
First, dependency risk carries over in a different form: 49.6\% of the skills with resolvable dependencies depend on a version with a known CVE, and unlike npm packages, which mostly become vulnerable after release, 78.3\% of them are \emph{born vulnerable}, already exposed at publication.
Second, few of these CVEs are reachable: agent input reaches the vulnerable function in only 3.0\% of the affected skills, whereas 21 of 22 sampled first-party flaws are exploited end to end in the sandbox.
Third, a live agent executes the injected payload in 48.5\% of the sampled flagged skills, but almost always in its own shell before the skill runs, and the payload reaches a flaw in the skill itself in only 5.0\%.
Hence, registries should prioritize first-party flaws over dependency alerts, and agent runtimes should not pass skill arguments through a shell.
\end{abstract}

%% file: sections/introduction.tex
\section{Introduction}
\label{sec:introduction}

Agent skills are emerging as a software-distribution ecosystem for extending agents based on large language models (LLMs)~\cite{jiang2026sok}.
A skill bundles executable code, third-party dependencies, installation logic, and natural-language instructions that tell an agent when and how to use it.
The \sysname registry~\cite{openclaw2024}, the largest public skill marketplace, hosted \numskills skills from \numauthors authors at our March 2026 snapshot, and 44.3\% of them contain executable code.
Because skills include code and dependencies, they inherit the security problems long studied in npm~\cite{zimmermann2019npm, decan2018impact, liu2022demystifying, zahan2022weak}, PyPI~\cite{alfadel2021empirical, vu2020typosquatting}, and Maven~\cite{zhang2023mitigating}.
A skill's runtime, however, changes \emph{who} invokes the code and \emph{how} its inputs are formed.
A developer writes and reviews the calls into a traditional package.
An LLM agent instead selects a skill, builds its arguments from a user request and external content, runs it, and processes its output.
Indirect prompt injection~\cite{greshake2023indirect, zhan2024injecagent} exploits this path by planting attacker instructions in content that the agent later loads.
A dangerous sink such as \texttt{eval()} or \texttt{os.system()} then executes that content as code on the user's machine.
Hence, an ordinary coding flaw or an outdated dependency that a benign author leaves in a skill becomes exploitable through the agent (\S\ref{sec:threat-model} traces an example).

Existing studies of skill security have two main limitations.
First, they target \emph{intentional} malice, such as backdoors, credential stealers, and planted injection payloads~\cite{liu2026maliciousagent, deng2026taming}, or a single attack vector~\cite{chen2026credleak}.
The \emph{unintentional} flaws that benign authors introduce, in dependencies and in their own code, have not been measured across a whole registry.
Second, registry-scale studies either report static alerts produced by regular expressions or LLM classifiers~\cite{liu2026agentwild}, or execute skills to confirm malicious behavior or credential leakage~\cite{liu2026maliciousagent, chen2026credleak}.
None checks whether an agent's input reaches a vulnerable dependency or first-party sink.
Work that does test this reachability analyzes one LLM-integrated application at a time~\cite{liu2024llmsmith, liu2025agentfuzz}.
Measuring these flaws at registry scale is hard for two reasons.
Static detectors over-approximate, so an alert must be confirmed by review before it is counted.
In addition, whether injected content arrives at a confirmed flaw depends on how the agent builds and runs the command, so reachability must be observed by execution.
Hence, the number of static alerts may differ widely from the number of flaws that an agent can trigger.

To this end, we conduct a registry-scale empirical study of the supply chain attack surface arising from unintentional vulnerabilities in agent skills, over all \numskills skills in \sysname.
We ask the following five research questions:

\rqbox{RQ1}{How prevalent are known vulnerabilities in skill dependencies, and how do they enter and persist?}
\rqbox{RQ2}{How prevalent are first-party code paths from agent input to command- or code-execution sinks?}
\rqbox{RQ3}{How often do risky install instructions reach the LLM, through installers and \texttt{SKILL.md}?}
\rqbox{RQ4}{How often do skills print credentials to stdout, where the agent's context captures them?}
\rqbox{RQ5}{Which flagged vulnerabilities can an agent actually reach and exploit?}

\noindent RQ1 and RQ2 measure the classical supply chain risks that skills inherit, RQ3 and RQ4 the attack vectors that the LLM runtime adds, and RQ5 whether an agent actually reaches the flagged flaws when it invokes the skill.
To answer them, we first screen every skill with \toolname, a high-recall static pipeline that combines software composition analysis (SCA) of dependencies, agent-aware static application security testing (SAST) of first-party code, and two agent-specific analyses of install instructions and stdout.
We then manually review all 3{,}428 cases flagged by the four main detectors, and every confirmed count comes from this manual census (\S\ref{sec:stage4}).
Finally, we measure reachability by execution: we reproduce 40 flagged cases in a sandbox harness and run 200 sampled flagged skills with a live LLM agent under indirect prompt injection (\S\ref{sec:repro-setup}).

Our study leads to three main findings.
First, classical supply chain risk carries over in a markedly different form (RQ1--RQ2).
Of the skills with resolvable manifests, 49.6\% depend on a version with a known CVE (Common Vulnerabilities and Exposures), and 78.3\% of them are \emph{born vulnerable}, that is, the CVE was already public when the skill was published, which is the reverse of the pattern in npm.
In first-party code, the census confirms 315 skills with an exploitable path from agent input to a command- or code-execution sink.
Second, the LLM runtime adds exposure that package ecosystems lack (RQ3--RQ4).
\texttt{SKILL.md}, which the LLM reads on every invocation, contains 6.6 to 22.8 times as many risky install instructions as the installers, whereas credentials printed to stdout are rare, with 22 confirmed leaks.
Third, and most importantly, static vulnerability prevalence poorly reflects runtime reachability (RQ5).
Agent input reaches the vulnerable function in only 3.0\% of the dependency-vulnerable skills, whereas 21 of the 22 first-party flaws reproduced in the sandbox harness are exploited end to end.
A live agent executes the injected payload in 48.5\% of the sampled flagged skills, but in almost all of them its own shell runs the payload before the skill starts, and the payload reaches the skill's own sink in only 5.0\%.
These findings suggest that registries should prioritize first-party flaws over dependency alerts, and that agent runtimes should not pass skill arguments through a shell.

\noindent This paper makes the following contributions:

\begin{itemize}
\item \textbf{Study.} To the best of our knowledge, we present the first registry-scale study of the supply chain attack surface arising from unintentional vulnerabilities in agent skills, covering dependencies, first-party code, install instructions, and stdout over all \numskills skills of \sysname.
\item \textbf{Carried-over risks.} We show that dependency risk carries over in a born-vulnerable form that authors rarely remediate, and we confirm 315 skills with an exploitable first-party path from agent input to a command- or code-execution sink.
\item \textbf{Agent-native risks.} We characterize two attack vectors that package analyses do not model, install instructions in \texttt{SKILL.md} and credentials printed to stdout, and show that risky install instructions appear mostly in documentation, where installer scans miss them, while confirmed credential leaks are rare.
\item \textbf{Reachability.} We show by execution that static prevalence poorly reflects what an agent can exploit: dependency CVEs are rarely reachable, first-party flaws are exploited in the sandbox harness, and most injected payloads run in the agent's own shell rather than in the skill.
\end{itemize}

%% file: sections/background.tex
\section{Background and Threat Model}
\label{sec:background}

\subsection{Agent Skills and Their Runtime}
\label{sec:bg-ecosystem}

An agent skill is a software package that extends an LLM-based agent with code and natural-language instructions.
It bundles a \texttt{SKILL.md} instruction file with optional \texttt{scripts/} code, an \texttt{install.sh} setup script, and \texttt{requirements.txt} or \texttt{package.json} manifests.
\texttt{SKILL.md} tells the agent when to invoke the skill and how to build the command that runs it.
Its usage block documents each command with placeholders, for example \texttt{python3 scripts/calc.py eval "<expression>"}.
The agent fills the placeholders with input derived from the user request or from external content.
A host framework such as Claude Code~\cite{anthropic2024claude} runs the resulting command on the user's machine with the user's privileges, captures its stdout and stderr, and inserts them into the LLM's conversation context.
The framework does not run the \texttt{install.sh}-family installer automatically~\cite{anthropic2025skills}, whereas the LLM reads \texttt{SKILL.md} on every invocation.
Authors publish skills to the registry without the lock files, dependency update bots, and continuous-integration checks with which mature package ecosystems keep known-vulnerable versions out.

\subsection{Threat Model}
\label{sec:threat-model}

\begin{figure}[t]
\centering
\includegraphics[width=0.5\figwidth]{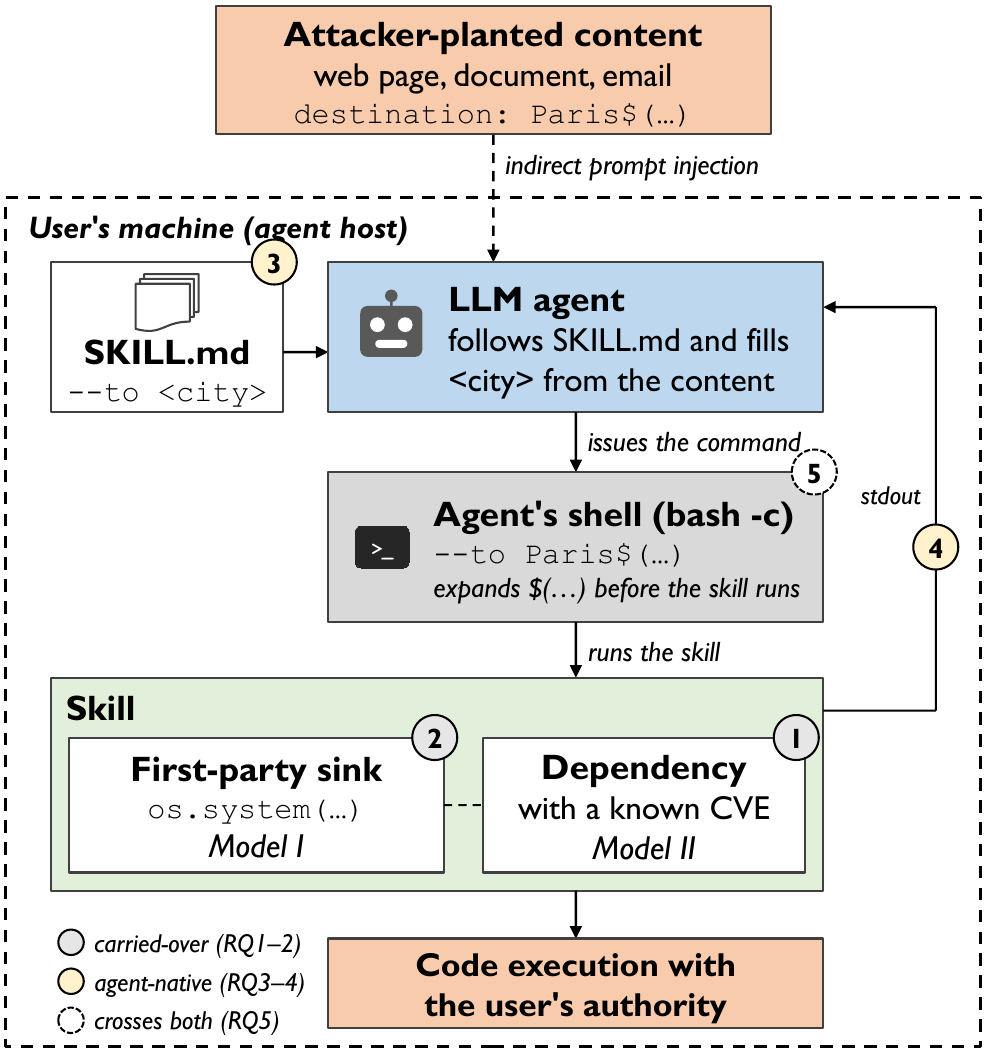}
\caption{Skill invocation and the five exposure points. \circled{1} and \circled{2} carry over from package ecosystems, and \circled{3} to \circled{5} are agent-native.}
\label{fig:threat}
\end{figure}

We study \emph{vulnerabilities}, weaknesses that skill authors introduce unintentionally.
Intentional malicious authorship, which we call \emph{malware}, is out of scope and has been studied concurrently~\cite{liu2026maliciousagent, guo2026malskillbench}.
The two need different methods: detecting malware requires evidence of malicious behavior, while vulnerabilities need taint analysis and dependency auditing of code written without malicious intent.

\noindent\textbf{Attacker model.}
The victim is a user who has installed a skill written by a benign author and runs it through an agent with a shell tool.
The attacker controls only external content that the agent processes, such as a web page, a document, or an email, and plants instructions in it (\emph{indirect prompt injection}~\cite{greshake2023indirect}).
The attacker cannot modify the skill, the registry, the user's request, or the LLM, but can read the skill's published code and \texttt{SKILL.md}.
The attacker's goal is to execute code or read credentials on the user's machine with the user's authority.

\noindent\textbf{Exposure points.}
We distinguish two mechanisms by which a benign author introduces a weakness.
\textbf{Model~I} is a first-party code vulnerability in the author's own scripts: command injection through \texttt{os.system} on unsanitized input, or code injection through \texttt{eval}/\texttt{exec} on external data.
In a skill, such a flaw escalates indirect prompt injection: when the injected content reaches the sink, it leads to code execution inside the skill.
\textbf{Model~II} is a vulnerable third-party dependency~\cite{ladisa2023taxonomy}: a declared dependency whose version has a known CVE.

Figure~\ref{fig:threat} traces the attack on a simplified running example, a travel skill (\S\ref{sec:results-rq2} shows the sink of a real one) whose \texttt{SKILL.md} tells the agent to fill a \code{<city>} argument from the content it processes.
The attacker places a shell command substitution in the destination field of the content, and the agent substitutes this field into the skill invocation.
The agent runs the skill through its own shell, which expands the substitution before the skill starts (\circled{5}), so this exposure point needs no flaw in the skill.
When the injected input reaches the skill, a Model~I sink gives the attacker code execution (\circled{2}), and a loaded Model~II dependency may expose a known CVE (\circled{1}).
Two further attack vectors have no counterpart in package managers.
Because the LLM reads \texttt{SKILL.md}, a risky install command that the file documents, such as \texttt{curl~\textbar~bash}, may be issued as a tool call although the installer itself is never run automatically (\circled{3}).
Because a skill's stdout re-enters the agent's context window, a credential that the skill prints is exposed to any later prompt injection in the same conversation (\circled{4}).
Exposure points \circled{1} and \circled{2} carry over from package ecosystems and are the subjects of RQ1 and RQ2, and \circled{3} and \circled{4} are agent-native and the subjects of RQ3 and RQ4.
RQ5 tests whether an agent actually reaches \circled{1} and \circled{2}, and also measures \circled{5}.

\subsection{Problem Formulation}
\label{sec:formulation}

We state the flaws that we measure in one formulation, which we reuse from taint analysis.
Following prior work~\cite{livshits2005finding,yamaguchi2014modeling}, let $\tau = (S, K, N)$ be a triple of \emph{sources}, \emph{sinks}, and \emph{sanitizers}.
A program has a \emph{$\tau$-flow} if data flows from a source $s \in S$ to a sink $k \in K$ without passing a sanitizer in $N$.
Table~\ref{tab:formulation} instantiates $\tau$ for three of the exposure points of Figure~\ref{fig:threat}.
A code vulnerability (\circled{2}) and a reachable dependency CVE (\circled{1}) share their sources, the inputs that the agent supplies to the skill, and differ in their sinks.
A stdout leak (\circled{4}) is a flow from a credential to stdout.
Install instructions (\circled{3}) are documented commands rather than flows, so we measure them by pattern matching (\S\ref{sec:tool}).

\begin{table*}[t]
\centering
\caption{Instantiations of the triple $\tau = (S, K, N)$ for the flaws that we measure.}
\label{tab:formulation}
\footnotesize
\setlength{\tabcolsep}{4pt}
\begin{tabular}{llll}
\toprule
\textbf{Flaw} & \textbf{Sources $S$} & \textbf{Sinks $K$} & \textbf{Sanitizers $N$} \\
\midrule
Code vulnerability (\circled{2}) & agent-supplied inputs & OS-command and code-evaluation calls & effective input checks \\
Reachable dependency CVE (\circled{1}) & agent-supplied inputs & the advisory's vulnerable function & effective input checks \\
Stdout leak (\circled{4}) & credential reads & writes to stdout & redaction calls \\
\bottomrule
\end{tabular}
\end{table*}

The classical setting assumes that the attacker supplies the source directly.
In a skill, the attacker controls only the content $c$ that the agent processes, and the agent builds the command $A(c)$ that invokes the skill from the user's request, \texttt{SKILL.md}, and $c$.
A flow from an agent-supplied input is therefore necessary but not sufficient for an attack on the skill, and we define exploitation at the level of the agent.

\begin{definition}[Exploitation by an agent]
\label{def:exploit}
A skill is \emph{exploited by an agent} $A$ if there is attacker-controlled content $c$ such that running $A(c)$ executes the attacker's payload.
The exploitation is on the \emph{skill-sink path} if the payload executes at the sink of a $\tau$-flow in the skill, and on the \emph{agent-shell path} if it executes when the agent's shell interprets $A(c)$, before the skill runs.
\end{definition}

The agent-shell path (\circled{5}) requires no flow in the skill.
RQ1--RQ4 measure the flows and the install patterns statically, and RQ5 measures exploitation by execution.

%% file: sections/studydesign.tex
\section{Study Design}
\label{sec:methodology}

\begin{figure*}[t]
\centering
\includegraphics[width=\figwidth]{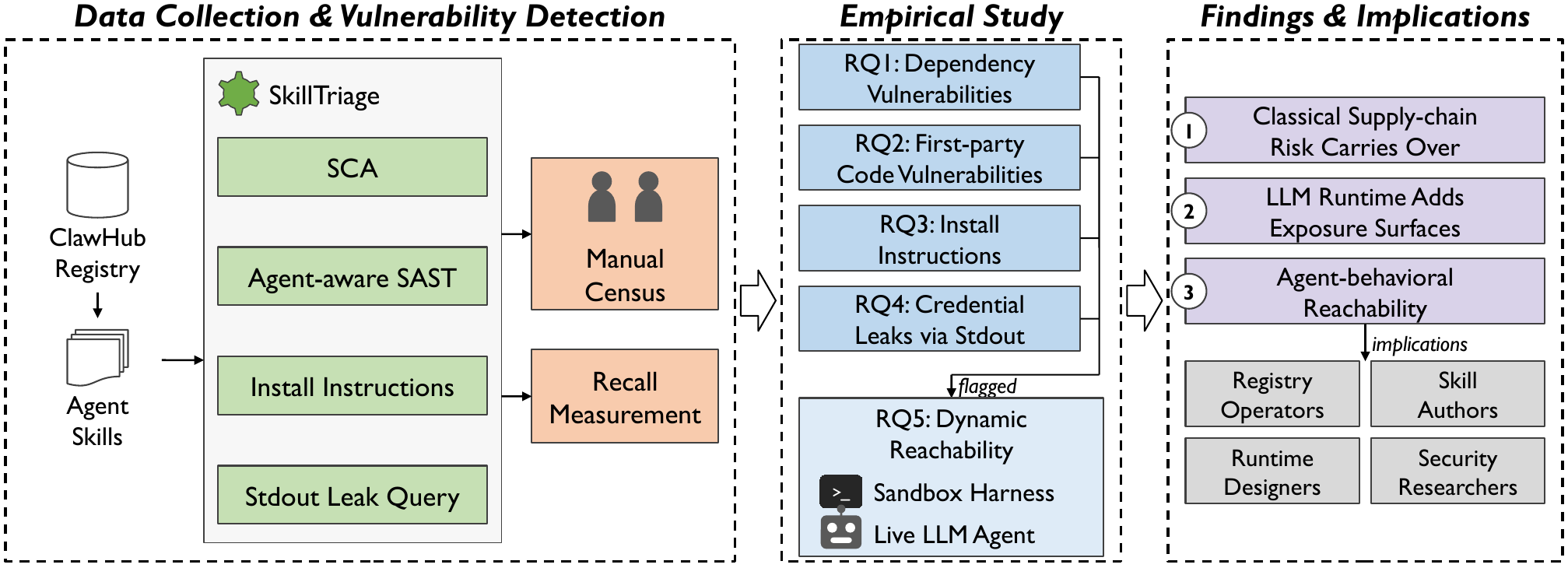}
\caption{Overview of the study.}
\label{fig:overview}
\end{figure*}

Figure~\ref{fig:overview} gives an overview of the study, which has three stages.
In the data collection and vulnerability detection stage, we collected the entire \sysname registry (\S\ref{sec:study-dataset}) and screened every skill with \toolname for the two vulnerability models and the two agent-native attack vectors of our threat model (\S\ref{sec:tool}).
Two authors then reviewed the flagged cases in a manual census (\S\ref{sec:stage4}), and we measured what the screen missed (\S\ref{sec:benchmark}).
In the empirical study stage, we answered RQ1--RQ4 on the confirmed cases and RQ5 by reproducing flagged cases in a sandbox harness and with a live LLM agent (\S\ref{sec:repro-setup}, \S\ref{sec:study}).
Finally, we derived findings and implications for registry operators, skill authors, runtime designers, and researchers (\S\ref{sec:discussion}).

\begin{figure*}[t]
\centering
\includegraphics[width=\figwidth]{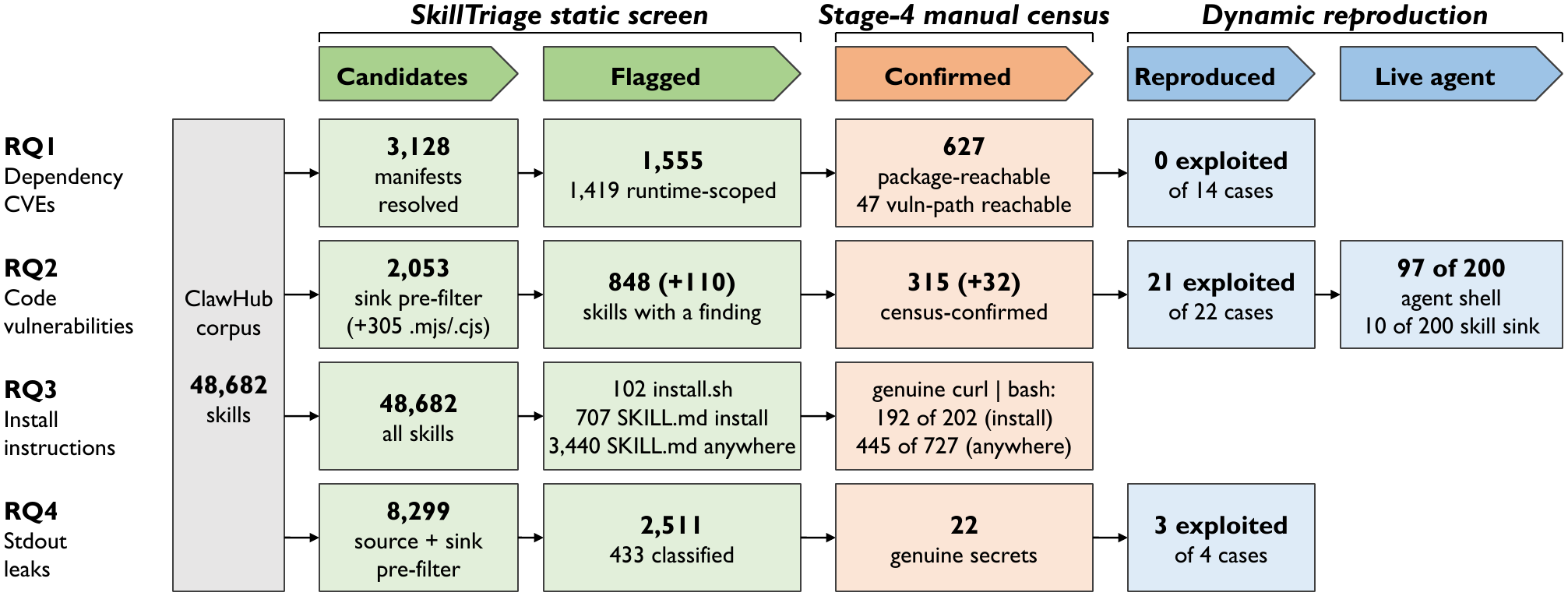}
\caption{Stages of every reported count, with the number of skills at each stage.}
\label{fig:stages}
\end{figure*}

Figure~\ref{fig:stages} shows the stages behind every count.
A skill is \emph{flagged} when a detector reports it, and \emph{confirmed} when the census judges the flagged case true under the detector's labeling criteria (\S\ref{sec:stage4}).
A flagged case is \emph{exploited end to end} when the sandbox harness executes a benign payload through it, and a skill is exploited by a live agent in the sense of Definition~\ref{def:exploit} (\S\ref{sec:repro-setup}).
In the RQ2 row of the figure, numbers in parentheses are the skills added when we extended the pre-filter to \texttt{.mjs} and \texttt{.cjs} files (\S\ref{sec:benchmark}).

\subsection{Dataset}
\label{sec:study-dataset}

\begin{figure}[t]
\centering
\includegraphics[width=0.40\figwidth]{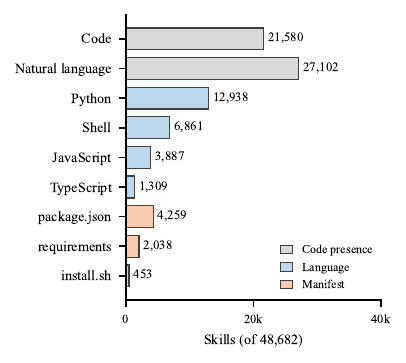}
\caption{Composition of the \sysname corpus.}
\label{fig:composition}
\end{figure}

We collected the full \sysname registry in March~2026 through its public Git mirror: \numskills skills from \numauthors authors.
We chose \sysname because it is the largest public skill registry.
As Figure~\ref{fig:composition} shows, 21{,}580 of the \numskills skills (44.3\%) contain executable code and the rest are pure natural language.
Python is the most common language, followed by shell and JavaScript, and a skill may mix them.
Most skills ship neither a \texttt{package.json} nor a \texttt{requirements.txt} and so declare no dependency.
This composition limits the detectors: the dependency scan covers only the skills with a manifest, and the code scan only the Python and JavaScript ones, so the denominators in RQ1 and RQ2 are these subsets, not the full corpus.
We counted every listed skill and did not merge near-identical copies, which \S\ref{sec:results-rq1} measures.
Besides the snapshot, RQ1 uses the publication dates in the registry metadata and a second fetch, six months later, of the current manifests and usage counters.

\subsection{Detection with \toolname}
\label{sec:tool}

\toolname runs one detector per vulnerability model and one per agent-native attack vector.
Each detector narrows the corpus to a flagged set that the census then reviews.

\noindent\textbf{Dependency vulnerabilities (SCA).}
\label{sec:tool-deps}
A skill has a \emph{dependency vulnerability} (Model~II) when a declared \texttt{(package, version)} falls in the affected range of an advisory in the Open Source Vulnerabilities database (OSV)~\cite{osv-database}.
We queried OSV because it is the only advisory database that covers PyPI, npm, and GitHub Advisories under one schema~\cite{imtiaz2021comparative}.
We ran \texttt{osv-scanner}~\cite{osv-scanner} on the 3{,}188 skills with a Python requirements file or an npm lock file.
For npm, the SCA covers only skills that include a lock file.
For Python, \texttt{osv-scanner} resolves a bare package name or a \texttt{>=} range to a concrete version at scan time, so a flagged version is not always one the author wrote.

\noindent\textbf{Code vulnerabilities (SAST).}
\label{sec:tool-code}
A skill has a \emph{code vulnerability} (Model~I) when it has a $\tau$-flow (\S\ref{sec:formulation}) that connects an agent-controlled source (\texttt{sys.argv}, \texttt{argparse}, \texttt{stdin}, or an environment variable the agent sets) to an OS-command (CWE-078) or code-evaluation (CWE-094) sink, as classified by the Common Weakness Enumeration (CWE)~\cite{mitre-cwe}.
We targeted only these two classes, because through them tainted agent input leads to code execution.
The detector has three stages.
\emph{Stage~1} is a regex pre-filter that keeps skills with at least one sink.
\emph{Stage~2} combines the results of two taint-analysis layers: CodeQL's built-in CWE-078/094 queries under a local threat model with these sources, and pattern queries for \texttt{eval}/\texttt{exec} on LLM-generated content and for shell strings built by f-string interpolation.
The pattern layer is syntactic, so it trades precision for recall.
\emph{Stage~3} filters out findings that match a known benign pattern: list-form \texttt{subprocess} without a shell, \texttt{exec} of a hardcoded sibling file, interpolation of constants such as \texttt{sys.executable}, and \texttt{DataFrame}/\texttt{model.eval}.

\begin{figure*}[t]
\centering
\includegraphics[width=\figwidth]{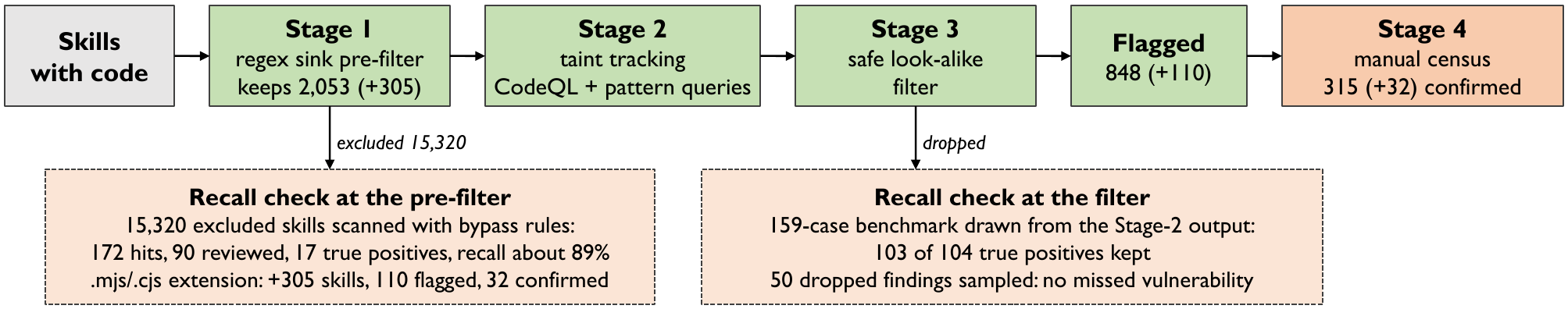}
\caption{Screening stages of the code-vulnerability detector and recall measurement of Stages~1 and~3 (\S\ref{sec:benchmark}).}
\label{fig:sast}
\end{figure*}

\noindent\textbf{Risky install instructions.}
\label{sec:tool-install}
\toolname scans \texttt{install.sh}-family installers and \texttt{SKILL.md} with twelve regular expressions for risky install patterns, such as \texttt{curl~\textbar~bash}, \texttt{eval} of fetched content, \texttt{chmod~777}, disabled SSL verification, and exporting a secret into the environment.
We scanned \texttt{SKILL.md} twice, on sections under an Install/Setup heading and on the whole file, because the precision of a match differs between the two (\S\ref{sec:results-rq3}).

\noindent\textbf{Stdout credential leaks.}
\label{sec:tool-stdout}
The stdout detector is a taint-tracking CodeQL query for the third instantiation of $\tau$, with credential sources and stdout-like sinks.
The sources are environment-variable reads, credential files such as \texttt{\textasciitilde/.ssh/} and \texttt{.env}, and \texttt{dotenv} APIs.
The sinks are \texttt{print}, loggers, and \texttt{console.*}, with calls to redaction-named functions treated as sanitizers.
Each skill with such a flow is assigned a severity level based on the credential evidence in the three lines before and four lines after the print (the $\pm$4 window, whose width \S\ref{sec:results-rq4} varies).
The STRONG tier covers SSH, AWS, \texttt{.env}, OAuth, and PEM credentials, the MEDIUM tier a named key variable, and the WEAK tier a generic environment variable.
A flow without credential evidence in the window is UNCLASSIFIED.

\subsection{Manual Census}
\label{sec:stage4}

\toolname serves only as a high-recall screen, and every confirmed count comes from the manual census, which Figures~\ref{fig:stages} and~\ref{fig:sast} show as Stage~4.

\noindent\textbf{Labeling criteria.} We fixed the criteria for each detector from its threat model before labeling.
For code vulnerabilities, a flagged skill is a true positive only if an agent-controlled source reaches a CWE-078/094 sink with no effective sanitizer, judged on the published code.
These criteria exclude operator- or deployment-set environment variables, list-form process calls, effective allowlists, and dead or test code.
For dependency CVEs, the census has two levels that follow prior work on dependency reachability.
A flagged dependency is \emph{package-reachable} when the skill imports the vulnerable package on a runtime path, which excludes development-only dependencies~\cite{pashchenko2018vulnerable} and declared packages that the skill never imports.
It is \emph{vulnerable-path-reachable} when agent input also reaches the advisory's vulnerable function~\cite{ponta2018beyond} in the package copy that the skill loads.
For the second level, we first recorded for each advisory its vulnerable function, the attacker input that triggers it, and any precondition, and then checked each skill against that record.
We excluded three advisories that report a malicious release rather than a vulnerable version.
For stdout leaks, a case counts only if a genuine secret value reaches stdout or stderr; a variable name, a masked preview, or a host name does not count.
For install instructions, a match counts only if the command would run when followed as documented.

\noindent\textbf{Annotators and inter-rater agreement.} Two authors, each with more than five years of experience in vulnerability and malware analysis, performed all labeling, which took about two person-months.
One author labeled every case of the four main detectors, that is, the flagged dependency CVEs, code vulnerabilities, and \texttt{curl~\textbar~bash} instructions, and the stdout flows with credential evidence (3{,}428 cases in all), so their confirmed counts are exact rather than estimated (Table~\ref{tab:stage4}).
The other install patterns and the stdout flows without credential evidence are labeled on samples (\S\ref{sec:results-rq3}, \S\ref{sec:results-rq4}).
Screen precision falls as the detection target becomes more contextual, from 95.0\% for install-section \texttt{curl~\textbar~bash} to 44.2\% for package-reachable dependency CVEs and 5.1\% for stdout credentials.
The other author independently labeled a seeded, stratified random subsample of 150 cases per detector.
At 150 cases, the 95\% Wilson interval of a precision estimate has a half-width of at most 0.08.
Cohen's $\kappa$ is 1.0 for stdout leaks and \texttt{curl~\textbar~bash}, 0.96 (95\% CI 0.91 to 1.0) for dependencies, and 0.62 (0.49 to 0.72) for code vulnerabilities.
For the vulnerable-path level, the second author wrote an independent trigger record and labeled a random sample of 50 skills, with $\kappa = 0.92$ (49 of 50 agree).
The two authors discussed every disagreement until consensus, and in a final round we tried to refute every true positive.
For code vulnerabilities, 17 of the 30 disagreements concern whether an environment variable is set by the agent or the operator.
None of the 30 became a counted true positive, since all 51 exploitable skills in the subsample were labeled true positive by both annotators.
To assess the reliability of the single-annotator census, the second author also labeled all 848 code-vulnerability cases independently.
The two labelings agree at $\kappa = 0.645$ (95\% CI 0.588 to 0.698).
The remaining disagreement concerns the same agent-versus-operator boundary, and confirmed counts use the census labels.

\begin{table}[t]
\centering
\caption{Manual census of every flagged case per detector; screen precision is confirmed divided by flagged.}
\label{tab:stage4}
\footnotesize
\setlength{\tabcolsep}{3pt}
\begin{tabular}{@{}>{\raggedright\arraybackslash}p{0.47\columnwidth}rrr@{}}
\toprule
\textbf{Detector} & \textbf{Flagged} & \textbf{Prec.} & \textbf{Confirmed} \\
\midrule
Dependency CVE, package-reachable (RQ1) & 1{,}419 & 44.2\% & 627 \\
Dependency CVE, vulnerable-path-reachable (RQ1) & 1{,}419 & 3.3\% & 47 \\
Code vulnerability (RQ2) & 848 & 37.1\% & 315 \\
Install-section \texttt{curl~\textbar~bash} (RQ3) & 202 & 95.0\% & 192 \\
Anywhere \texttt{curl~\textbar~bash} (RQ3) & 727 & 61.2\% & 445 \\
Stdout leak, credential-classified (RQ4) & 433 & 5.1\% & 22 \\
\bottomrule
\end{tabular}
\end{table}

\noindent\textbf{Separating vulnerabilities from malware.}
\label{sec:vuln-malware-rubric}
Our scope is unintentional flaws, so we separated them from malice by observable signals rather than inferred intent.
A skill is labeled malware if it shows at least two of eight signals: obfuscation that hides the skill's purpose, hardcoded exfiltration to an endpoint unrelated to the declared purpose, access to credential stores unrelated to the declared purpose, secrecy directives in its instruction file, hidden Unicode or HTML-comment directives, packer artifacts, a decoy-payload split, or impersonation of a legitimate tool.
Among the 315 confirmed code vulnerabilities, 13 skills match one of the five automatically detectable signals, but manual review confirms none of them.
Even if every automatic match counted as malware, the count would fall only from 315 to 302.
Under a stricter one-signal rule, our review reclassifies none of them.
One flagged false positive meets the two-signal rule; after our report, the registry removed it.

\noindent\textbf{Alternatives to manual review.} We manually reviewed every flagged case because no automatic confirmation reached a precision that could substitute for it.
On the 848 code-vulnerability cases, a static filter that removes list-form \texttt{subprocess} calls and interpolation of system constants is nearly lossless (41.9\% precision at 99.7\% recall).
However, every LLM-based confirmation trades recall for precision: an LLM judge over the sink's $\pm k$-line context reaches 60.0\% precision at 90.5\% recall, and a whole-function judge reaches 66.5\% precision at 63.5\% recall.
On the 433 stdout cases the same judge reaches 81\% to 89\% precision but misses 5 or 6 of the 22 confirmed leaks.
Hence, we labeled every flagged case by hand.

\subsection{Recall of the Code-Vulnerability Screen}
\label{sec:benchmark}

A vulnerability that \toolname misses never enters the census, so the confirmed counts are lower bounds and \toolname needs high recall.
As Figure~\ref{fig:sast} shows, \toolname can discard a vulnerability at two stages, and we measured the recall of both.
\emph{Stage~3 filter.} We built a benchmark of 159 Python sink locations, sampled per rule from the Stage-2 output: 105 built-in command-injection, 15 built-in code-injection, 24 f-string, and 15 \texttt{eval}/\texttt{exec} findings.
Two authors labeled the benchmark with 81.8\% agreement.
Of its 104 true positives, Stage~3 keeps 103 (99.0\%), and none of 50 randomly sampled findings that Stage~3 filtered out is a missed vulnerability (one-sided 95\% upper bound 5.8\%).
\emph{Stage~1 pre-filter.} We scanned the 15{,}320 code-bearing skills that it excludes, using complementary rules that match the sink variants it misses.
These rules produced 172 matches, and we reviewed 90 of them.
Extrapolating the per-stratum rates gives about 39 missed vulnerable skills (24 to 64), a recall of about 89.0\% (83.0\% to 93.0\%).
A uniform sample of 200 excluded skills contains one missed vulnerability (one-sided 95\% upper bound 2.4\%), the same skill that the complementary rules find.
Almost all misses have one cause: the JavaScript pre-filter did not scan \texttt{.mjs} and \texttt{.cjs} files.
Hence, we extended the pre-filter to these files and manually reviewed the 305 affected skills.
Of them, 110 are flagged and 32 are confirmed, which leaves about five estimated misses.
We report these 32 skills separately: the headline count remains the 315 skills confirmed in the original 848-skill census, on which the analyses of RQ2 and the samples of RQ5 are based, and the extended total is 347.
\emph{Language coverage.} The SAST covers only Python and JavaScript, while 6{,}861 skills (14.1\%) also include shell scripts.
Two shell scans flag 254 skills.
Reviewing 145 of them confirms 8 more exploitable skills, which is a lower bound.
Skills that pass Stage~1 but have no Stage-2 finding were not sampled, so we cannot bound the number of flows that CodeQL misses.
For stdout leaks, \S\ref{sec:results-rq4} estimates the leaks among the UNCLASSIFIED flows.

\subsection{Dynamic Reproduction}
\label{sec:repro-setup}

\noindent\textbf{Sandbox harness.}
We reproduced 40 flagged cases: an initial 20 that include the case study of RQ2, and a 20-case expansion chosen to cover each CWE class and both sink families.
For each case, a per-skill harness injected a benign payload built from canary credentials and a deterministic marker.
The harness counts a case as exploited end to end when the marker reaches the exfiltration endpoint, stdout, or stderr.

\noindent\textbf{Live agent.}
In a pilot study, we first examined 18 harness skills in depth with the agent described below, and five of them also with \texttt{gpt-4o}, to observe how the LLM handles injected values.
We then drew a seeded random sample of 200 skills, stratified by language (88 Python and 112 JavaScript), from the 827 flagged skills that are not in the harness set.
The agent was \texttt{gpt-4o-mini} with a \texttt{run\_command} tool that executes through \texttt{bash -c}, under naive and explicit indirect injection, with three trials each.
A \emph{naive} injection places the payload in a field of the external content, and an \emph{explicit} injection adds an instruction that tells the agent to pass the field unchanged.
A trial fires when the marker of the payload reaches the drop file.
To separate exploitation through the agent's shell from exploitation at the skill's own sink, a paired control re-ran every exploited skill in the same container under two conditions.
The \emph{shell condition} kept the \texttt{bash -c} tool.
The \emph{execve condition} replaced it with a \texttt{run\_program(program, args)} tool that executes an argument vector and rejects direct shell calls.
Hence, under the execve condition, a command substitution reaches the skill as literal text.
The two conditions separate the paths of Definition~\ref{def:exploit}: exploitation that only the shell condition reproduces is on the agent-shell path, and exploitation that persists under the execve condition is on the skill-sink path.
The \emph{skill-sink rate} is the share of sampled skills exploited through the skill-sink path.

\noindent\textbf{Ethical considerations.}
All dynamic execution ran in an isolated sandbox with canary credentials and blocked egress, so no real credential or user data was read or sent.
We analyzed only published source and accessed no live deployment.
We responsibly disclosed the confirmed vulnerabilities to the registry operators.

%% file: sections/study.tex
\section{Empirical Results}
\label{sec:study}

\subsection{RQ1: Dependency Vulnerabilities}
\label{sec:results-rq1}

Skill authors publish without the safeguards of mature package ecosystems (\S\ref{sec:bg-ecosystem}).
This RQ asks how many skills depend on a version with a known CVE, whether the CVE was already public when the skill was published, and which practices introduce such versions and leave them unfixed.

\begin{figure*}[t]
\centering
\begin{minipage}{\figwidth}
\begin{minipage}[t]{0.48\linewidth}
\centering
\includegraphics[width=\linewidth]{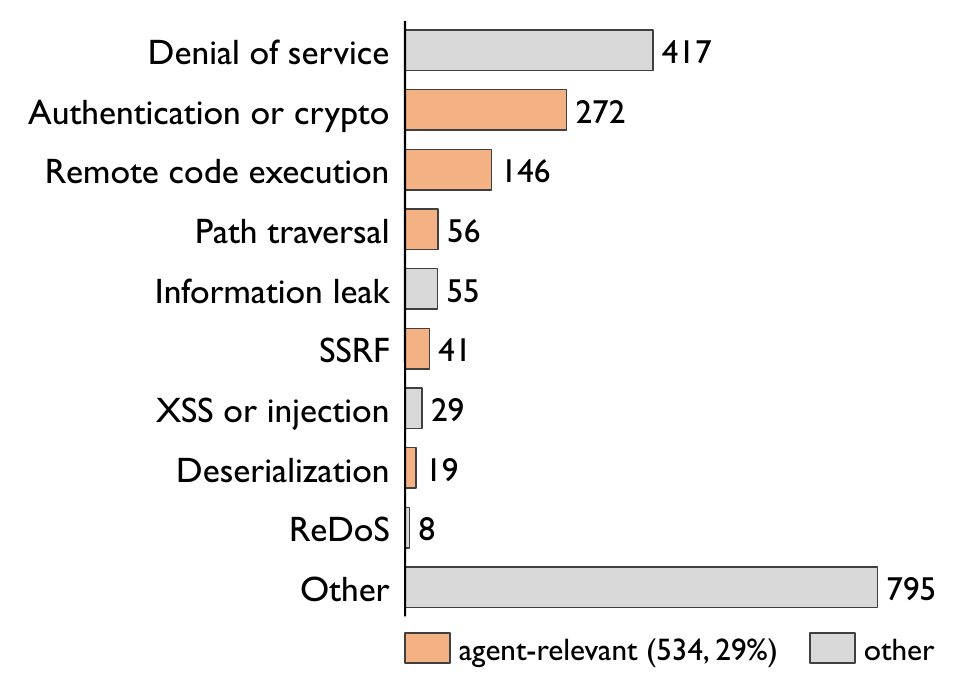}
\caption{Unique dependency advisories by type.}
\label{fig:advisory-types}
\end{minipage}\hfill
\begin{minipage}[t]{0.48\linewidth}
\centering
\includegraphics[width=\linewidth]{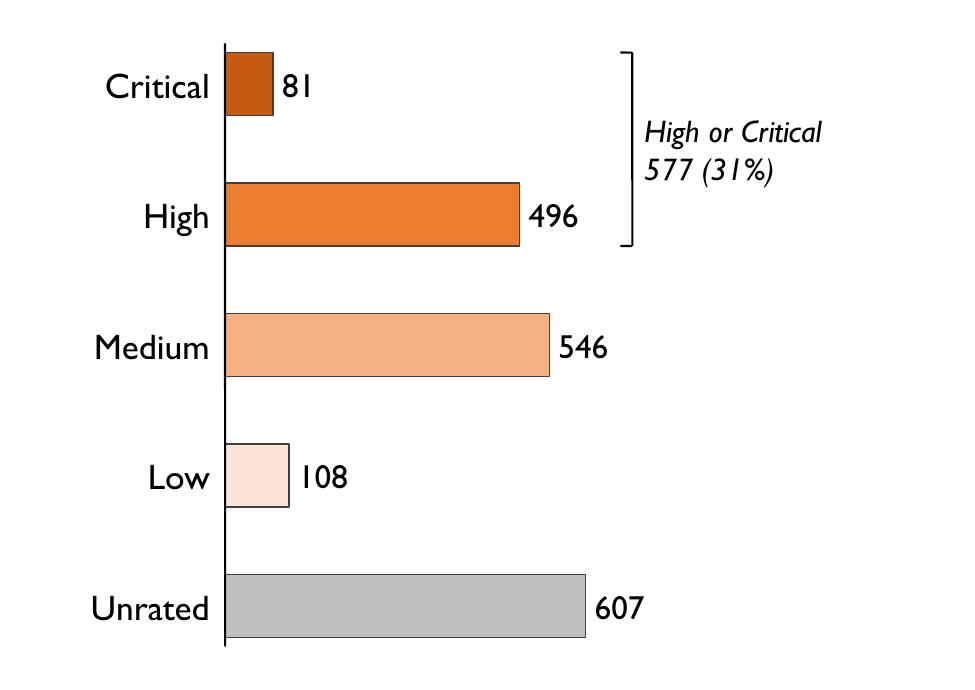}
\caption{Unique dependency advisories by severity.}
\label{fig:advisory-severity}
\end{minipage}
\end{minipage}
\end{figure*}

\noindent\textbf{Prevalence.}
The SCA resolves concrete versions for 3{,}128 of the 3{,}188 skills it scans, and about half of those (49.6\%) depend on a version with a known CVE: 1{,}555 skills in all (3.2\% of the corpus), affected by 1{,}838 advisory records (1{,}474 distinct vulnerabilities once GitHub (GHSA) and PyPI (PYSEC) advisory aliases are merged) (Figure~\ref{fig:advisory-types}).
However, only 22.8\% of the flagged versions are pinned by the author, and the rest come from transitive lock-file entries or from bare package names that the resolver pinned.
Restricting to exact pins and lock files leaves the prevalence almost unchanged (50.7\%).

\begin{figure*}[t]
\centering
\begin{minipage}{\figwidth}
\begin{minipage}[t]{0.40\linewidth}
\centering
\includegraphics[width=\linewidth]{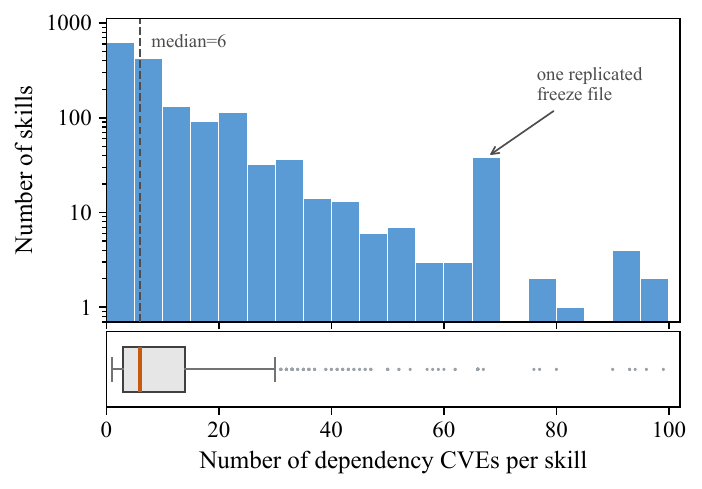}
\caption{Per-skill dependency advisory counts, as (package, advisory) entries.}
\label{fig:vuln-distribution}
\end{minipage}\hfill
\begin{minipage}[t]{0.57\linewidth}
\centering
\includegraphics[width=\linewidth]{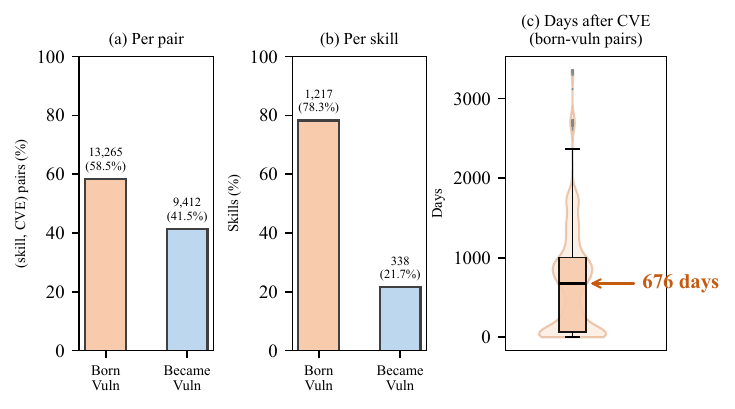}
\caption{Born-vulnerable versus became-vulnerable dependencies, by pair, by skill, and by disclosure-to-publication lag.}
\label{fig:born-vuln}
\end{minipage}
\end{minipage}
\end{figure*}

As Figure~\ref{fig:vuln-distribution} shows, the per-skill distribution is heavy-tailed: 45.7\% of affected skills have one to five advisories, and 34 skills have more than 50.
Nearly a third of the advisories are severe: 577 of the 1{,}838 (31\%, Figure~\ref{fig:advisory-severity}) are rated High or Critical.

\begin{figure*}[t]
\centering
\includegraphics[width=\figwidth]{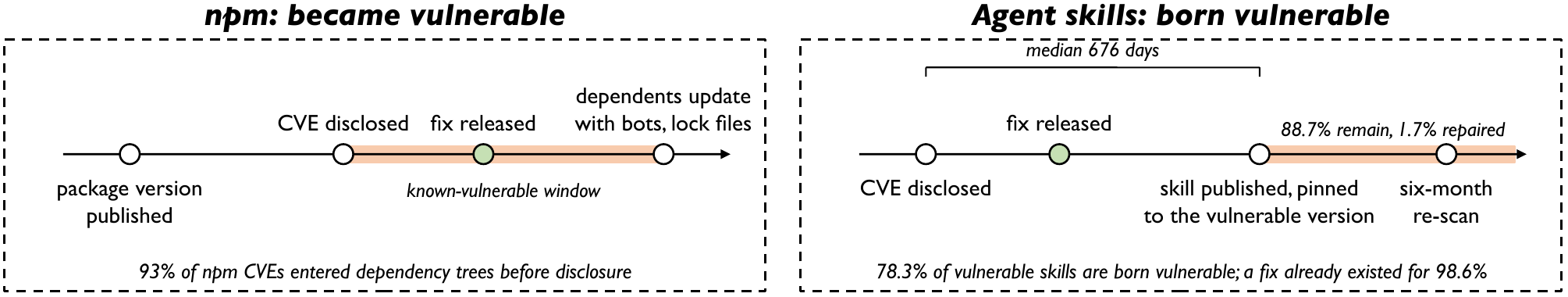}
\caption{Why skills are born vulnerable: in npm a dependency becomes vulnerable after it is installed, while a skill is published already pinned to a version with a public CVE.}
\label{fig:timeline}
\end{figure*}

\noindent\textbf{Born vulnerable.}
In npm, 93\% of CVEs entered dependency trees before disclosure, so packages \emph{became} vulnerable later~\cite{liu2022demystifying}.
As Figure~\ref{fig:timeline} shows, the order is reversed in skills, where the vulnerable version is already pinned at publication.
To measure this order, we compared each skill's first-publication date with each CVE's disclosure date (Figure~\ref{fig:born-vuln}).
In total, 58.5\% of (skill, CVE) pairs are born vulnerable, since the author published the skill after the CVE was already public.
At the skill level, 78.3\% of vulnerable skills (1{,}217 of 1{,}555) have at least one CVE known at publication, and 57.1\% (435 of 762) on pinned versions alone.
For born-vulnerable pairs, the CVE had been public for a median of 676 days when the skill was published.
This exposure was avoidable, because a fixed release existed at publication for 98.6\% of the (manifest entry, advisory) pairs whose advisory was already public.
Being born vulnerable also changes remediation.
For a became-vulnerable dependency, a bot can propagate the later upstream fix.
For a born-vulnerable pin, however, the fix already existed and the author did not adopt it, so no later propagation event removes the vulnerability.

\begin{findingbox}
\textbf{Finding 1:} 1{,}555 skills (3.2\% of the corpus) depend on a version with a known CVE, 49.6\% of the skills whose dependencies resolve to concrete versions.
Unlike npm dependencies, which become vulnerable after installation, 78.3\% are born vulnerable (57.1\% when only author-pinned versions are counted).
That is, they pin a version whose CVE was already public at publication, a median of 676 days earlier.
\end{findingbox}

\noindent\textbf{Origin and persistence of vulnerable versions.}
We traced every vulnerable exact Python pin to its file and author.
Most of these pins, 1{,}158 of 1{,}350 (85.8\%), come from \texttt{pip freeze} dumps of an operating-system Python environment that five authors committed, and one 127-line freeze file, copied into 40 skills by one author, contains 1{,}092 of them.
This copying is pervasive: near-duplicate clustering (MinHash over \texttt{SKILL.md} and the scripts, Jaccard $\geq 0.85$) groups the corpus into 2{,}538 clusters of two or more skills, 97\% of them genuine duplicates in a 100-cluster sample (two authors, $\kappa = 1.0$) and a fifth copied across different authors, so a vulnerable template reappears unchanged in many copies.
The mechanism differs by ecosystem.
For Python, a vulnerable exact pin was the newest release at publication in 14.6\% of cases, against 43.1\% for clean pins, which indicates that exposure results from stale pinning.
For npm, a vulnerable dependency was the newest release as often as a clean one (53.6\% versus 58.0\%), which suggests that the advisory usually appeared after the lock file was written.
Dependency hygiene does not prevent either case: 69.6\% of npm skills (2{,}966 of 4{,}259) include no lock file, against over 90\% lock-file adoption in mainstream npm~\cite{zimmermann2019npm}.

After publication, skills are rarely repaired.
Six months after the snapshot, we re-fetched the current manifests of the 1{,}279 still-listed dependency-vulnerable skills and re-scanned them with the same analyzer version.
Of the 16{,}153 (skill, advisory) pairs from March whose status the re-scan could determine, 88.7\% still appear in the current manifest, and the author had upgraded or removed the vulnerable dependency in only 1.7\%.
However, authors do release updates, since 55\% of the dependency-vulnerable skills had released a new version by then, against 43\% of all skills.
Hence, publishing an update is common, but removing the vulnerable dependency is rare.

\begin{findingbox}
\textbf{Finding 2:} Most vulnerable exact Python pins (85.8\%) come from \texttt{pip freeze} dumps that five authors committed.
Six months later, 88.7\% of the advisories remain and authors had removed the vulnerable dependency in only 1.7\%, though 55\% of these skills had released a new version.
\end{findingbox}

\subsection{RQ2: First-Party Code Vulnerabilities}
\label{sec:results-rq2}

Unlike a library function, a skill script receives arguments that the agent builds from user requests and external content.
This RQ measures how many skills pass such arguments to a command- or code-execution sink without an effective sanitizer, using the SAST detector and the census of \S\ref{sec:methodology}.

\begin{table*}[t]
\centering
\caption{Census of the 848 flagged skills, each counted under the layer, category, and language of its representative finding; Python and JS cells give flagged / confirmed skills.}
\label{tab:code-vulns}
\footnotesize
\begin{tabular}{llrrrrr}
\toprule
\textbf{Layer} & \textbf{Category} & \textbf{Python} & \textbf{JS} & \textbf{Flagged} & \textbf{Confirmed} & \textbf{Precision} \\
\midrule
\multirow{2}{*}{Built-in} & Command injection & 213 / 63 & 428 / 201 & 641 & 264 & 41.2\% \\
 & Code injection & 18 / 7 & 15 / 5 & 33 & 12 & 36.4\% \\
\midrule
\multirow{2}{*}{Custom pattern} & Shell string interpolation & 43 / 10 & 84 / 15 & 127 & 25 & 19.7\% \\
 & \texttt{eval}/\texttt{exec} of non-literal & 37 / 12 & 10 / 2 & 47 & 14 & 29.8\% \\
\midrule
\multicolumn{2}{l}{\textbf{Total}} & \textbf{311 / 92} & \textbf{537 / 223} & \textbf{848} & \textbf{315} & \textbf{37.1\%} \\
\bottomrule
\end{tabular}
\end{table*}

\noindent\textbf{Prevalence.}
As Table~\ref{tab:code-vulns} shows, 848 skills (1.74\%) are flagged with at least one path from agent input to a command- or code-execution sink, 13 of them in both languages.
Command injection dominates, and JavaScript accounts for most flagged skills, mainly through \texttt{child\_process.execSync} with template-literal arguments.
The census confirms 315 of the 848 as exploitable (37.1\%).
Extending the pre-filter to \texttt{.mjs} and \texttt{.cjs} files later confirms 32 more (\S\ref{sec:benchmark}).
The other 533 are mostly analysis false positives (388), taint paths whose source is outside the agent threat model (111), or dead, test, or example code (30).

\begin{findingbox}
\textbf{Finding 3:} The census confirms 315 skills with an exploitable path from agent input to a command- or code-execution sink, 37.1\% of the 848 flagged, mostly command injection.\end{findingbox}

\noindent\textbf{Case study: evaluating LLM-generated code.}
The most direct chain passes attacker-influenced LLM output to the interpreter (Listing~\ref{lst:vibe-reading}).
The skill sends e-book excerpts to an LLM, which returns a Python scanning function that the skill executes with full built-ins.
An attacker hides an instruction in the book text, and the LLM emits the payload alongside the expected function.
The payload then runs with access to the LLM API key, the user's SSH credentials, and the secrets that other skills store under the shared agent home directory.
Every step is the skill's intended behavior, and a retry loop that feeds errors back to the LLM allows the payload three further attempts.
Thus, the attacker's content inherits the skill's ambient authority~\cite{hardy1988confused}.

\begin{lstlisting}[float=tb, caption={Vulnerable chapter scanning in the \texttt{drbillwang/vibe-reading} skill.},
  label={lst:vibe-reading}]
# Send book excerpt to an LLM, then exec the code it returns
response = self.client.generate_content(prompt)
scan_code = json.loads(response)['code']
exec(scan_code,
     {'__builtins__': __builtins__,  # full builtins exposed
      're': re, 'json': json})
\end{lstlisting}

\noindent\textbf{Case study: interpolating arguments into a shell command.}
The more common chain builds a shell string from agent-extracted arguments.
This enterprise travel skill, which appears in the corpus as four near-identical copies, concatenates the agent's positional arguments, unsanitized, into \texttt{os.system} at nine call sites (Listing~\ref{lst:travel-api}), and its \texttt{SKILL.md} documents commands whose arguments are city names, dates, and passenger details that the agent takes from the user's request.
An attacker-controlled travel-approval email can contain a command substitution disguised as a destination city, which the shell then expands.
However, the commands that \texttt{SKILL.md} documents call the per-task scripts directly and bypass this dispatcher, and in our live-agent runs the agent's own shell expanded the substitution before any skill code ran (\S\ref{sec:results-rq5}).
The skill stores authentication tokens and passenger identity numbers on disk, so the injected command can exfiltrate both.
The LLM provides only incidental defense, such as normalizing a city name, and an explicit instruction in the injected content overrides it.

\begin{lstlisting}[float=tb, caption={Unsanitized argument interpolation in the \texttt{gaogao605/fenbeitong-travel} dispatcher.}, label={lst:travel-api}]
args = sys.argv[3:]   # city, date, passenger details from the agent
if action == "search":
    os.system(f"python3 {SCRIPTS_DIR}/flight_search.py {' '.join(args)}")
elif action == "order":
    os.system(f"python3 {SCRIPTS_DIR}/flight_order.py {' '.join(args)}")
# ... nine os.system dispatches, each joins args into a shell string
\end{lstlisting}

\noindent\textbf{Injection paths documented by authors.}
Some authors document the injection path themselves.
In their own \texttt{SKILL.md}, 81 of the 844 flagged skills with a \texttt{SKILL.md} (9.6\%) instruct the agent to pass input derived from user or external content to the vulnerable script.
This rate is 2.3 times the rate among unflagged skills (4.1\%).
In 16 confirmed cases, \texttt{SKILL.md} also documents the dangerous operation itself, for example by asking the LLM to generate the code that the script executes.

\noindent\textbf{Structural factors.}
Repeating the structural analysis of prior work~\cite{liu2026agentwild} with being flagged by \toolname as the outcome, we find one protective factor: skills from \emph{multi-skill authors}, those with three or more skills, are less likely to be flagged.
The effect remains significant in a multivariable logistic regression over the 22{,}301 script-bearing skills that controls for code size, dependency count, \texttt{install.sh}, and language (adjusted odds ratio 0.79, 95\% CI 0.72 to 0.87).

\begin{findingbox}
\textbf{Finding 4:} Some authors document the injection path themselves: 81 flagged skills (9.6\%) instruct the agent in \texttt{SKILL.md} to pass user or external content to the vulnerable script, 2.3 times the rate among unflagged skills.
Skills from authors with three or more skills are less likely to be flagged (adjusted odds ratio 0.79).
\end{findingbox}

\subsection{RQ3: Install Instructions in Installers and Documentation}
\label{sec:results-rq3}

A skill has two install locations: the \texttt{install.sh}-family installer, which the host framework does not run automatically, and \texttt{SKILL.md}, whose commands enter the LLM's context and may be issued as tool calls (\S\ref{sec:threat-model}).
This RQ compares how often risky install instructions appear in the two locations, and how many of the pattern matches are genuine.
On raw regex counts, the twelve risk patterns flag 102 skills in \texttt{install.sh} (0.21\%), 707 in a \texttt{SKILL.md} install section (1.45\%), and 3{,}440 anywhere in \texttt{SKILL.md} (7.07\%, Table~\ref{tab:install-patterns}).

\begin{table}[t]
\centering
\caption{Install-instruction risk by location, as raw regex matches and as genuine instructions after per-pattern precision weighting (95\% CIs in text).}
\label{tab:install-patterns}
\footnotesize
\setlength{\tabcolsep}{4pt}
\begin{tabular}{lrr}
\toprule
\textbf{Location} & \textbf{Raw} & \textbf{Genuine} \\
\midrule
\texttt{install.sh}            & 102     & 40 \\
\texttt{SKILL.md} install sec. & 707     & 266 \\
\texttt{SKILL.md} anywhere     & 3{,}440 & 910 \\
\bottomrule
\end{tabular}
\end{table}
However, a regex match is not necessarily a risky instruction.
Hence, we manually labeled every \texttt{curl~\textbar~bash} match in \texttt{SKILL.md} and a stratified sample of up to 30 matches per pattern and location (381 cases, $\kappa = 0.97$).
Precision depends on the pattern.
\texttt{sudo} installs, installs from Git or a URL, and \texttt{curl} to a file are 75\% to 100\% genuine in \texttt{SKILL.md}, while \texttt{chmod~777}, \texttt{eval} of fetched content, and secret export are almost never genuine (0 of 93).
The secret-export pattern alone accounts for 2{,}232 of the 3{,}440 matches anywhere in \texttt{SKILL.md}, because it matches security warnings and placeholders.
Weighting each location by its patterns' precision gives about 40 skills with a genuine risky pattern in \texttt{install.sh} (95\% CI 33 to 48), 266 in a \texttt{SKILL.md} install section (259 to 271), and 910 anywhere in \texttt{SKILL.md} (878 to 940).
Thus, on raw regex matches the ratio of \texttt{SKILL.md} to \texttt{install.sh} is 7 for install sections and 34 for the whole file, and on genuine instructions it is 6.6 (95\% CI 5.5 to 8.2) and 22.8 (18.8 to 28.0).

Precision also depends on where the instruction appears in \texttt{SKILL.md}.
For \texttt{curl~\textbar~bash}, 192 of 202 matches under an Install/Setup heading are genuine (95.0\%), against 253 of 525 elsewhere (48.2\%).
We treat these \texttt{SKILL.md} instructions as \emph{LLM-context exposure} rather than execution, because a documented command runs only if the LLM issues it.

\begin{findingbox}
\textbf{Finding 5:} Risky install instructions appear mostly in documentation: \texttt{SKILL.md} contains 6.6 to 22.8 times as many genuine ones as \texttt{install.sh} (an estimated 266 skills in install sections and 910 in the whole file, against 40).
They do not run automatically but enter the LLM's context on every invocation, so a scan of installers alone misses most of them.
\end{findingbox}

\subsection{RQ4: Credential Leaks through Stdout}
\label{sec:results-rq4}

A credential that a skill prints to stdout enters the LLM's context (\S\ref{sec:threat-model}), where any later prompt injection can read it, unlike a traditional log file that needs filesystem access.
Chen~\etal~\cite{chen2026credleak} report that this pattern accounts for 73.5\% of credential-leak issues in another registry.
In this RQ, we measure it on \sysname.

Our stdout query flags 2{,}511 skills (\toolleakratepct of the corpus).
Table~\ref{tab:rq4-severity} gives their severity tiers: 433 have credential evidence in the $\pm$4-line window, and 2{,}075 are UNCLASSIFIED, a taint flow with no credential pattern near the print.
The census of all 433 confirms only 22 that print a genuine secret value (5.1\%), because the query cannot distinguish a secret from a host, a path, a public client identifier, a credential name, or a masked preview.

\begin{table}[t]
\centering
\caption{Highest severity tier per flagged skill, from credential evidence in the $\pm$4-line window.}
\label{tab:rq4-severity}
\footnotesize
\begin{tabular}{lrr}
\toprule
\textbf{Severity} & \textbf{Skills} & \textbf{\% of ecosystem} \\
\midrule
STRONG & 149 & 0.31\% \\
MEDIUM & 44 & 0.09\% \\
WEAK & 240 & 0.49\% \\
UNCLASSIFIED & 2{,}075 & 4.26\% \\
Unreadable & 3 & 0.01\% \\
\midrule
\textbf{Total} & \textbf{2{,}511} & \textbf{5.16\%} \\
\bottomrule
\end{tabular}
\end{table}

\begin{figure*}[t]
\centering
\includegraphics[width=\figwidth]{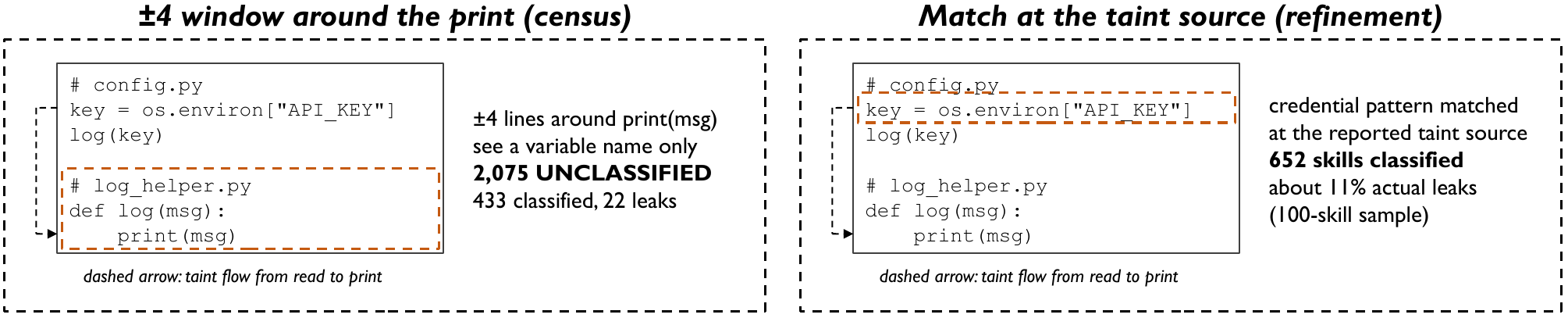}
\caption{Matching the credential pattern at the taint source rather than around the \texttt{print} recovers a leak whose credential is read in another function.}
\label{fig:anchor}
\end{figure*}

The 22 leaks are a lower bound, and the limit is where the window is placed rather than how wide it is.
The $\pm$8, $\pm$16, and whole-function windows keep all 22 but add 273 to 597 skills, of which only 1 of 34 sampled is an actual leak.
In contrast, a uniform sample of 100 UNCLASSIFIED skills contains 6 actual leaks (95\% CI 2.8\% to 12.5\%), which extrapolates to about 125 among the 2{,}075, and in 5 of the 6 the secret is read in another function or file, so no window around the print can contain it.
As Figure~\ref{fig:anchor} shows, matching the same severity patterns at the credential source reported by the taint query classifies 652 of the UNCLASSIFIED skills, and about 11\% of a 100-skill sample of them are actual leaks (95\% CI 6.3\% to 18.6\%).\footnote{This source-based variant uses a later, stricter revision of the query, so its population is not identical to the 2{,}075. Hence, we report it as a refinement of the detector rather than as an addition to the 22 confirmed leaks.}
For comparison, Chen~\etal~\cite{chen2026credleak} confirm 352 leaks in 17{,}022 skills by execution, which also finds secrets that appear only at run time.

\begin{findingbox}
\textbf{Finding 6:} A credential printed to stdout enters the agent's context, but actual leaks are rare.
The static query over-flags heavily: the census confirms 22 leaks, and a sample of the UNCLASSIFIED flows suggests about 125 more.
\end{findingbox}

\subsection{RQ5: Reachability under Agent Invocation}
\label{sec:results-rq5}
\label{sec:dynamic-reproduction}

RQ1--RQ4 count flaws that static analysis and code review find, but an attacker can exploit a flaw only if the agent passes attacker content to it.
This RQ measures reachability by execution: the sandbox harness for first-party flaws and dependency CVEs, a source check for compound skills, and a live agent for the flagged code-vulnerability skills.

\begin{table}[t]
\centering
\caption{Harness reproduction of 40 flagged cases.}
\label{tab:reproduction-summary}
\footnotesize
\begin{tabular}{lrr}
\toprule
\textbf{Class} & \textbf{Cases} & \textbf{Exploited} \\
\midrule
CWE-094 (\texttt{eval}/\texttt{exec}) & 14 & 14 \\
CWE-078 (command injection)           &  7 &  7 \\
CWE-089 (SQL injection)               &  1 &  0 \\
Stdout credential leak                &  4 &  3 \\
Dependency CVE (SCA)                  & 14 &  0 \\
\midrule
\textbf{Total}                        & \textbf{40} & \textbf{24} \\
\bottomrule
\end{tabular}
\end{table}
\begin{figure*}[t]
\centering
\includegraphics[width=\figwidth]{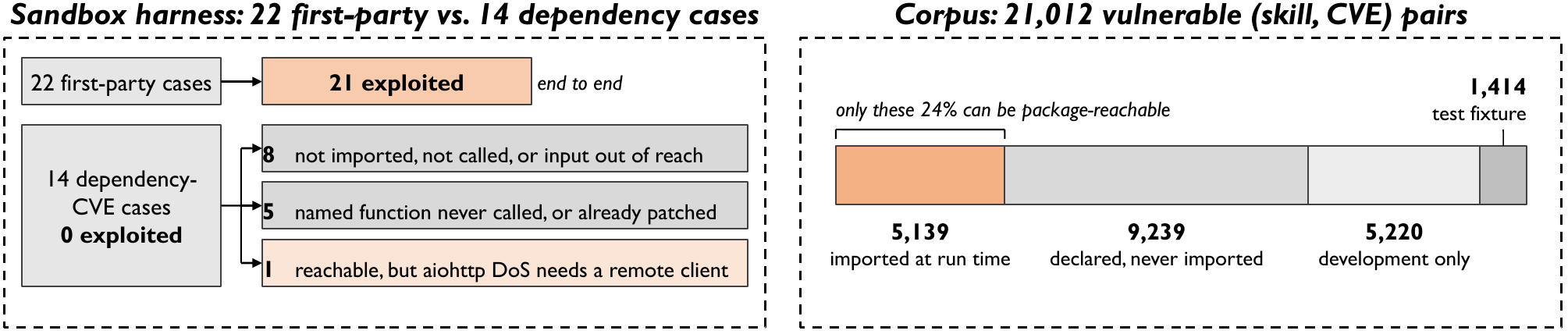}
\caption{Why dependency CVEs are rarely reachable: outcomes of the 14 dependency cases in the harness (left), and the 21{,}012 vulnerable (skill, CVE) pairs by how the skill uses the package (right).}
\label{fig:depreach}
\end{figure*}

\noindent\textbf{First-party flaws versus dependency CVEs.}
As Figure~\ref{fig:depreach} shows, the harness outcomes are asymmetric.
In the harness, 21 of the 22 first-party cases are exploited end to end, while none of the 14 dependency-CVE cases is (Table~\ref{tab:reproduction-summary}).
The one first-party failure is a flagged SQL injection outside our sink set, and its identifier check rejects the payload at the entry point.
A first-party flaw is reachable because the skill receives agent-constructed arguments as input.
A dependency CVE is usually not reachable, because its vulnerable function is one call site in an imported library and is rarely on the path that the agent triggers.
Of the 14 dependency cases, eight fail because the library is not imported, its vulnerable function is not called (\eg, \texttt{torch}), or agent input cannot reach the vulnerable argument (\eg, \texttt{requests} skills that pin their API host).
Checked against the function that each advisory names, five more are not reachable either.
Three \texttt{pillow} cases are flagged for an advisory on \texttt{ImageMath.eval}, which none of the skills calls, an \texttt{lxml} case is flagged for a denial of service in \texttt{iterwalk}, which the skill does not use, and in a \texttt{qs} case the skill loads a patched copy, while the flagged version is a nested copy that only another library uses.
The remaining \texttt{aiohttp} case is reachable, but its server-side denial of service needs a remote client rather than agent input.
With $n = 14$, we report the direction of this asymmetry, not a precise rate.

At corpus scale, most flagged dependency CVEs are out of the agent's reach, as prior work reports for open-source dependencies~\cite{pashchenko2018vulnerable,ponta2018beyond}.
First, only 5{,}139 of the 21{,}012 vulnerable (skill, CVE) pairs (24\%) are in a package that the skill imports at run time.
The rest are development-only (5{,}220), test fixtures (1{,}414), or declared but never imported (9{,}239).
Second, the census confirms a reachable vulnerable path in only 47 of the 1{,}555 flagged skills (3.0\%).
Third, the 47 skills depend on a few advisories.
Three advisories account for 28 of them (59.6\%), that is, a \texttt{.netrc} credential leak in \texttt{requests}, a stack overflow in \texttt{yaml}, and an infinite loop in \texttt{PyPDF2}.
The 15 \texttt{requests} skills leak only if a \texttt{.netrc} file exists.
Only 5 of the 47 skills (10.6\%) depend on an advisory of an agent-relevant type, that is, remote code execution (RCE), deserialization, server-side request forgery (SSRF), path traversal, or authentication and cryptographic flaws, while 29\% of all 1{,}838 advisory records (534, Figure~\ref{fig:advisory-types}) are of these types.
The largest named group of the rest is denial of service (417), which needs a server role that the agent never takes.
Hence, a vulnerable version in a manifest rarely becomes a flaw that an agent can trigger.

Four harness cases include a defense, three custom \texttt{eval} sandboxes and one \texttt{open()} allowlist, and all four are bypassed, but only with an input crafted for the specific sandbox.
In the live-agent runs below, a sandbox of the same type blocked the payload that the agent passed on, so these defenses make an attack harder but do not remove the path.
Every harness exploit enters through the input that the agent passes to the skill, either as an argument or as content that the argument names.

\noindent\textbf{Compound skills.}
We find 36 skills (0.07\%) that have both a confirmed code vulnerability and a known dependency CVE, and 2 of them are among the 47 with a reachable dependency path.
A compound skill is the only case in which a first-party path could make a dependency CVE reachable.
We checked all 36 in source and found no such chain.
In 34, attacker input never reaches the vulnerable function, and in the other 2 it reaches a vulnerable YAML parser that only enables a crash before the injection sink.\footnote{For example, \texttt{rikisann/\allowbreak video-proof} parses an agent-supplied YAML file with \texttt{yaml}~2.8.2 and passes a field to a shell.}
In every case, the first-party flaw already grants code execution.

\begin{findingbox}
\textbf{Finding 7:} Once input reaches the skill, first-party flaws are exploitable, but dependency CVEs rarely are.
In the harness, 21 of 22 first-party flaws are exploited end to end and none of the 14 dependency CVEs is.
At corpus scale, only 47 of the 1{,}555 dependency-vulnerable skills (3.0\%) have a reachable vulnerable path, and no compound skill chains a first-party flaw to a dependency CVE.
\end{findingbox}

\begin{figure*}[t]
\centering
\includegraphics[width=\figwidth]{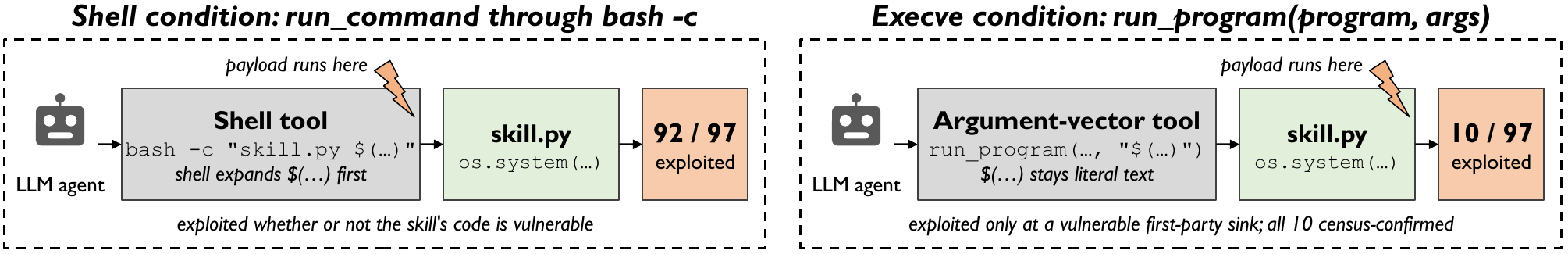}
\caption{Paired control on the 97 live-agent exploits. Under the shell condition the payload executes whether or not the skill is vulnerable, but under the execve condition only at a vulnerable first-party sink.}
\label{fig:paths}
\end{figure*}

\noindent\textbf{Live-agent exploitation.}
In 97 of the 200 sampled flagged skills (48.5\%, Wilson 95\% CI 42 to 55), the live agent executes the injected payload.
Another 18 reach the sink, but a missing dependency prevents the payload from executing, so a complete installation would raise the rate to at most 57.5\%.
Of the remaining skills, 18 are blocked by an in-skill guard, and 67 (33.5\%) are not reachable by the agent.
However, in 95 of the 97 exploited skills, the agent placed the attacker's command substitution \texttt{\$(...)} unquoted in its own command line, so the agent's shell expanded it before the skill ran.
This path does not depend on the skill's code: of the 97 skills, 41 are census-confirmed and 56 are census false positives.
The paired control separates the two paths (Figure~\ref{fig:paths}).
The shell condition reproduces 92 of the 97 exploits (94.8\%, 95\% CI 88.5 to 97.8), while under the execve condition the payload reaches the skill's own sink in only 10 (10.3\%, 95\% CI 5.7 to 17.9).
Paired by skill, 10 are exploited under both conditions, 82 only under the shell one, none only under the execve one, and 5 under neither (exact McNemar $p < 10^{-24}$).
All 10 skills exploited under execve are census-confirmed, and none of the 56 census false positives is exploited under it, so the census and the dynamic result agree.
Hence, the skill-sink rate is 10 of 200 (5.0\%, 95\% CI 2.7 to 9.0).
This is a lower bound, since the 18 skills blocked by a missing dependency were not re-run.

\noindent\textbf{Census labels versus live-agent outcomes.}
Table~\ref{tab:census-vs-agent} cross-tabulates the census label of each sampled skill with its live-agent outcome.
None of the 130 census-negative skills is exploited at its own sink (one-sided 95\% upper bound 2.3\%), so the census missed no skill-sink exploit in this sample.
In contrast, of the 70 census-positive skills, only 10 are exploited at their own sink.
The other 60 are either exploited only through the agent's shell or stop before the sink, at the agent, at a missing dependency, or at an in-skill guard.
Therefore, the census rate of 37.1\% measures what code review can confirm, while the skill-sink rate of 5.0\% measures what one agent exercises.
The gap between the two rates comes from the agent and the runtime rather than from the code.

\begin{table}[t]
\centering
\caption{Census label versus live-agent outcome for the 200 sampled skills.}
\label{tab:census-vs-agent}
\footnotesize
\setlength{\tabcolsep}{3pt}
\begin{tabular}{@{}>{\raggedright\arraybackslash}p{0.50\columnwidth}rrr@{}}
\toprule
\textbf{Live-agent outcome} & \textbf{Census pos.} & \textbf{Census neg.} & \textbf{Total} \\
\midrule
Payload executes at the skill's own sink & 10 & 0 & 10 \\
Payload executes, but not at the skill's sink & 31 & 56 & 87 \\
Reaches the sink, dependency missing & 13 & 5 & 18 \\
Blocked by an in-skill guard & 3 & 15 & 18 \\
Not reachable by the agent & 13 & 54 & 67 \\
\midrule
Total & 70 & 130 & 200 \\
\bottomrule
\end{tabular}
\end{table}

\noindent\textbf{Argument sanitization by the LLM.}
The LLM sanitizes arguments inconsistently, and its behavior depends on how the injected value looks.
In the 18-skill pilot study, the LLM rewrote natural-language fields such as a destination city under naive injection, and the payload executed in none of 9 such trials.
In contrast, it passed opaque arguments such as URLs and dates unchanged, and those skills were exploited under naive injection.
However, an explicit instruction in the injected content overrode the rewriting.
In the 200-skill sample, 21 skills that are not exploited under naive injection are exploited once the instruction is explicit.
\texttt{gpt-4o} refused values that were recognizably code even under explicit instruction, so a stronger model blocks more obvious payloads but not values that look like ordinary arguments.
Hence, a defense that relies on the LLM to sanitize arguments fails exactly on the inputs that a skill expects, and the check should be placed at the skill's entry point.

\begin{findingbox}
\textbf{Finding 8:} Injected payloads mostly run in the agent's own shell rather than in the skill.
A live agent executes the payload in 48.5\% of the sampled flagged skills, but delivers it to the skill's own sink in only 5.0\%, and the LLM does not reliably sanitize the arguments it passes.
\end{findingbox}

\noindent\textbf{Overlap and usage of the confirmed skills.}
The three confirmed risk types cover almost disjoint sets of skills.
In all, 381 skills have at least one confirmed risk: a reachable dependency CVE (47), an exploitable code vulnerability (315), or an actual stdout leak (22).
Only 3 of them have two of the three types, and none has all three, so a scanner for a single risk type would miss the risks of the other two.
These skills remain in use: six months after the snapshot, 83\% of them remain listed, and together they have been downloaded 529{,}181 times.
On the registry's usage counters, fetched at the six-month mark, their median download count is slightly above the corpus (1{,}154 versus 1{,}041), and against a baseline of unflagged skills that also contain code they are more downloaded (Mann--Whitney rank-biserial $r = 0.14$, $p < 10^{-4}$) and over-represented among the top 10\% most-downloaded (13.6\% versus 10.0\%).
Download counts are inflated by mirrors, but installs, the cleaner usage signal, agree.
Among the 264 confirmed-exploitable skills outside the double-labeled subsample, 46\% read a secret and perform network I/O in the same code, so one injected command can exfiltrate a credential.

%% file: sections/discussion.tex
\section{Discussion}
\label{sec:discussion}

\subsection{Implications}
\label{sec:implications}

Our findings lead to three lessons, which the actions below apply.
First, the risk comes from the few flaws that an agent reaches rather than from the many dependency CVEs, so triage should rank alerts by reachability.
Second, the agent rather than a developer decides which command runs on which input, so defenses should be placed at the skill's entry point and in the agent's shell.
Third, 78.3\% of vulnerable skills are already vulnerable at publication and six months later authors had removed the vulnerable dependency for only 1.7\% of the advisories, so the check should run at publication.

\noindent\textbf{Registry operators.}
\ding{182}~Rank first-party taint alerts above dependency alerts, and drop development-only and never-imported dependencies before notifying authors (Finding~7).
\ding{183}~Check pinned versions against OSV at publication, and require an npm lock file (Findings~1 and~2).
\ding{184}~Check \texttt{SKILL.md} automatically, using only the install patterns with high precision (Finding~5).
\ding{185}~Publish advisories for the confirmed vulnerable skills that are still listed, ranked by reachability (\S\ref{sec:results-rq5} and Finding~7).
\ding{186}~Group near-duplicate skills, so one fix or alert covers the cluster (\S\ref{sec:results-rq1}).

\noindent\textbf{Skill authors.}
\ding{182}~Declare direct dependencies rather than include a frozen environment, and check them before publishing (Finding~2).
\ding{183}~Validate agent-supplied input at the skill's entry point, where every exploit in our sandbox harness enters (Finding~7).
\ding{184}~Redact secrets before printing them, and do not document a command that passes external content into a shell sink (Findings~4 and~6).

\noindent\textbf{Runtime designers.}
\ding{182}~Run a skill's documented command through a tool that executes an argument vector rather than a shell, which lowers the share of the 97 exploited skills in which the payload executes from 94.8\% to 10.3\% without changing any skill (Finding~8).
\ding{183}~Redact stdout and stderr before they re-enter the context (Finding~6).
\ding{184}~Run skills under least privilege, since 46\% of the examined confirmed-exploitable skills read a secret and access the network in the same process (\S\ref{sec:results-rq5}).

\noindent\textbf{Security researchers.}
\ding{182}~Model the agent's selection and invocation path when assessing exploitability (Findings~7 and~8).
\ding{183}~Build dedicated detectors for the two agent-native attack vectors, install instructions in documentation and credential leakage from stdout into the context, and for the agent's shell (Findings~5, 6, and~8), which off-the-shelf SCA and SAST tools are not designed to cover.

\subsection{Threats to Validity}
\label{sec:limitations}

\noindent\textbf{Internal validity.}
Our confirmed counts depend on expert judgment and on what the screen flags.
Two authors labeled under criteria fixed before labeling, with $\kappa$ from 0.62 to 1.0, and no disagreement entered a confirmed count (\S\ref{sec:stage4}).
A vulnerability that \toolname never flags cannot enter the census, so we measured the screen's recall and extended it where it missed (\S\ref{sec:benchmark}).
Coverage is partial: the SCA needs a resolvable manifest, the SAST covers only Python and JavaScript, and some JavaScript CodeQL databases failed to build.
Hence, every confirmed count is a lower bound.

\noindent\textbf{External validity.}
We measure one registry at one point in time, so our rates may not generalize to other registries such as skillsmp.com, although concurrent studies of other registries~\cite{liu2026agentwild, liu2026maliciousagent, chen2026credleak} report rates of the same order.
The live-agent run uses GPT models (\texttt{gpt-4o-mini}, and \texttt{gpt-4o} on five pilot skills~\cite{openai2024gpt4o}) rather than the models of production agent hosts, with three unseeded trials per variant.
Hence, the exact exploitation rates are specific to our test harness.
However, the two reachability paths depend on conditions that hold beyond our harness.
The agent-shell path needs only an unsanitized shell tool, which current command-line agents expose, and the skill-sink path depends on the skill's code rather than on the LLM.

\noindent\textbf{Construct validity.}
A static alert does not imply an exploitable flaw, so we separate a confirmed flaw, which the census judges on the code, from an exploited one, which a sandbox harness or a live agent exploits end to end.
These two settings approximate production agent hosts but do not reproduce every one of them, so the reachability rates describe only what these settings exercise.

%% file: sections/relatedwork.tex
\section{Related Work}
\label{sec:related}

\subsection{Security of Agent Skills}
Several concurrent studies measure security in this ecosystem, mostly via regex or LLM classification rather than taint analysis and SCA.
Liu~\etal~\cite{liu2026agentwild} apply a dangerous-pattern taxonomy.
Three studies target intentional attacks: a follow-up study~\cite{liu2026maliciousagent} behaviorally confirms \emph{malicious} skills and reports that 84.2\% of malicious behavior resides in \texttt{SKILL.md} rather than code, which supports our focus on \texttt{SKILL.md} as an attack vector, Saha~\etal~\cite{saha2026skillmd} study \emph{intentional} semantic attacks delivered through \texttt{SKILL.md}, and Schmotz~\etal~\cite{schmotz2026skill} measure agent exposure to skill-file attacks.
Others study a single attack vector or a smaller sample: Chen~\etal~\cite{chen2026credleak} present the first study of credential leakage in skills, Zhu~\etal~\cite{zhu2026skillclone} study skill clones, Guo~\etal~\cite{guo2026skillprobe} audit 2{,}500 skills with cooperating LLM agents, and SkillFortify~\cite{bhardwaj2026skillfortify} formally analyzes skill supply chain properties and evaluates them on 540 skills.
Chen~\etal's study is closest in scope, and we study the same leakage on a different registry (\S\ref{sec:results-rq4}).
None of these studies enumerates dependency CVEs or confirms an interprocedural flow from agent input to a sink.
In contrast, we measure the attack surface arising from \emph{unintentional} vulnerabilities in dependencies and first-party code, confirm the flagged cases of the four main detectors by hand, and separate flagged flaws from those that an agent exploits (\S\ref{sec:dynamic-reproduction}).

\subsection{Software Supply Chain Security}
Supply chain risk in open-source ecosystems is well studied: attack taxonomies~\cite{ohm2020backstabber,ladisa2023taxonomy}, the concentration of npm risk in a few maintainers~\cite{zimmermann2019npm}, vulnerability propagation through dependency trees~\cite{decan2018impact,liu2022demystifying,hu2024empirical}, Python package vulnerabilities~\cite{alfadel2021empirical}, and weaknesses in package signing infrastructure~\cite{zahan2022weak}.
Package-manager attacks such as typosquatting and dependency confusion~\cite{birsan2021dependency,duan2021measuring,vu2020typosquatting} are intentional attacks and are outside our unintentional-vulnerability scope.
Whether to pin or float a dependency is itself contested: He~\etal~\cite{he2025pinning} show that pinning alone does not defend against supply chain attacks, and Rahman~\etal~\cite{rahman2025pinning} find that floating reduces outdated and vulnerable dependencies.
We adopt this prevalence-and-hygiene perspective, but applied directly it misestimates the risk in two ways.
First, the presence of a vulnerability no longer implies exposure.
Whether a vulnerable dependency or first-party sink matters depends on whether an LLM agent's input path reaches it, so we separate flagged flaws from those that an agent exploits (\S\ref{sec:dynamic-reproduction}).
Second, the LLM runtime adds attack vectors that this line of work did not model, such as install instructions in documentation and credential leakage from stdout into the agent's context.
In addition, the hygiene comparison gives the opposite result to that in mature registries.
Skills lack the lock files, dependency update bots, and continuous integration that keep known-vulnerable packages out of those registries.
Hence, most dependency-vulnerable skills are born vulnerable, that is, exposed at publication rather than by a later disclosure (\S\ref{sec:results-rq1}).

\subsection{LLM Agent Security}
Greshake~\etal~\cite{greshake2023indirect} introduced indirect prompt injection.
Subsequent work benchmarks or sandboxes agent behavior~\cite{zhan2024injecagent,debenedetti2024agentdojo,ruan2024toolemu,xi2025agents,zhan2025adaptive}.
These studies work at the level of prompts and agent behavior, rather than the code and dependencies of the tools agents invoke.
Closer to us, LLMSmith~\cite{liu2024llmsmith} and AgentFuzz~\cite{liu2025agentfuzz} analyze the code of \emph{individual} LLM-integrated applications with static analysis and fuzzing.
We instead measure the skill ecosystem at scale by static analysis and confirm reachability on samples by execution.
Studies of Model Context Protocol (MCP) servers~\cite{hou2025mcp,mehedi2025mcp} examine a different extension mechanism.
Runtime defenses such as SkillGuard~\cite{pan2026skillguard} and sandboxing are orthogonal to our study: they restrict or isolate what a skill may do, whereas we quantify the latent flaws in its code and dependencies.

%% file: sections/conclusion.tex
\section{Conclusion}
\label{sec:conclusion}

This paper presents a registry-scale study of the supply chain attack surface arising from unintentional vulnerabilities in agent skills, combining a manual census of every flagged case with reachability measured by execution.
Classical dependency risk carries over in a born-vulnerable form that later maintenance rarely repairs.
The LLM runtime adds exposure that package ecosystems lack, through install instructions in \texttt{SKILL.md} and credentials printed to stdout.
Reachability depends on how the agent invokes a skill.
In the harness, almost all first-party flaws are exploited end to end, while dependency CVEs almost never are.
A live agent executes the injected payload mostly in its own shell.
Therefore, securing this ecosystem needs reachability-aware triage at the registry and boundary controls in the runtime.

%% file: main.bbl
\begin{thebibliography}{10}

\bibitem{alfadel2021empirical}
Mahmoud Alfadel, Diego~Elias Costa, and Emad Shihab.
\newblock Empirical analysis of security vulnerabilities in {Python} packages.
\newblock {\em Empirical Software Engineering}, 28(3):59, 2023.
\newblock \href {https://doi.org/10.1007/s10664-022-10278-4}
  {\path{doi:10.1007/s10664-022-10278-4}}.

\bibitem{anthropic2024claude}
{Anthropic}.
\newblock Claude code: {AI}-powered coding assistant.
\newblock \url{https://code.claude.com/docs}, 2024.
\newblock Accessed: May 2026.

\bibitem{anthropic2025skills}
{Anthropic}.
\newblock Extend {Claude} with skills: Claude {Code} documentation.
\newblock \url{https://code.claude.com/docs/en/skills}, 2025.
\newblock Accessed: May 2026.

\bibitem{bhardwaj2026skillfortify}
Varun~Pratap Bhardwaj.
\newblock Formal analysis and supply chain security for agentic {AI} skills.
\newblock {\em arXiv preprint arXiv:2603.00195}, 2026.

\bibitem{birsan2021dependency}
Alex Birsan.
\newblock Dependency confusion: How i hacked into {Apple}, {Microsoft} and
  dozens of other companies.
\newblock
  \url{https://medium.com/@alex.birsan/dependency-confusion-4a5d60fec610},
  2021.
\newblock Accessed: May 2026.

\bibitem{chen2026credleak}
Zhihao Chen, Ying Zhang, Yi~Liu, Gelei Deng, Yuekang Li, Yanjun Zhang, Jianting
  Ning, Leo~Yu Zhang, Lei Ma, and Zhiqiang Li.
\newblock How your credentials are leaked by {LLM} agent skills: An empirical
  study.
\newblock In {\em Proceedings of the 41st IEEE/ACM International Conference on
  Automated Software Engineering}, 2026.
\newblock \href {https://doi.org/10.1145/3832783.3834373}
  {\path{doi:10.1145/3832783.3834373}}.

\bibitem{debenedetti2024agentdojo}
Edoardo Debenedetti, Jie Zhang, Mislav Balunovic, Luca Beurer-Kellner, Marc
  Fischer, and Florian Tram{\`e}r.
\newblock {AgentDojo}: A dynamic environment to evaluate prompt injection
  attacks and defenses for {LLM} agents.
\newblock In {\em Advances in Neural Information Processing Systems 37 (NeurIPS
  2024)}, pages 82895--82920, 2024.
\newblock \href {https://doi.org/10.52202/079017-2636}
  {\path{doi:10.52202/079017-2636}}.

\bibitem{decan2018impact}
Alexandre Decan, Tom Mens, and Eleni Constantinou.
\newblock On the impact of security vulnerabilities in the npm package
  dependency network.
\newblock In {\em Proceedings of the 15th International Conference on Mining
  Software Repositories}, pages 181--191, 2018.
\newblock \href {https://doi.org/10.1145/3196398.3196401}
  {\path{doi:10.1145/3196398.3196401}}.

\bibitem{deng2026taming}
Xinhao Deng, Yixiang Zhang, Jiaqing Wu, Jiaqi Bai, Sibo Yi, Zhuoheng Zou, Yue
  Xiao, Rennai Qiu, Jianan Ma, Jialuo Chen, Xiaohu Du, Xiaofang Yang, Shiwen
  Cui, Changhua Meng, Weiqiang Wang, Jiaxing Song, Ke~Xu, and Qi~Li.
\newblock Taming openclaw: Security analysis and mitigation of autonomous {LLM}
  agent threats.
\newblock {\em arXiv preprint arXiv:2603.11619}, 2026.

\bibitem{duan2021measuring}
Ruian Duan, Omar Alrawi, Ranjita~Pai Kasturi, Ryan Elder, Brendan
  Saltaformaggio, and Wenke Lee.
\newblock Towards measuring supply chain attacks on package managers for
  interpreted languages.
\newblock In {\em 28th Annual Network and Distributed System Security
  Symposium, {NDSS} 2021, virtually, February 21-25, 2021}. The Internet
  Society, 2021.
\newblock \href {https://doi.org/10.14722/ndss.2021.23055}
  {\path{doi:10.14722/ndss.2021.23055}}.

\bibitem{osv-database}
{Google}.
\newblock {OSV}: Open source vulnerabilities database.
\newblock \url{https://osv.dev/}, 2026.
\newblock Accessed: May 2026.

\bibitem{osv-scanner}
{Google Open Source Security Team}.
\newblock {OSV-Scanner}: Vulnerability scanner for open source.
\newblock \url{https://github.com/google/osv-scanner}, 2026.
\newblock Accessed: May 2026.

\bibitem{greshake2023indirect}
Kai Greshake, Sahar Abdelnabi, Shailesh Mishra, Christoph Endres, Thorsten
  Holz, and Mario Fritz.
\newblock Not what you've signed up for: Compromising real-world
  {LLM}-integrated applications with indirect prompt injection.
\newblock In {\em Proceedings of the 16th ACM Workshop on Artificial
  Intelligence and Security}, pages 79--90, 2023.
\newblock \href {https://doi.org/10.1145/3605764.3623985}
  {\path{doi:10.1145/3605764.3623985}}.

\bibitem{guo2026malskillbench}
Wenbo Guo, Wei Zeng, Chengwei Liu, Xiaojun Jia, Yijia Xu, Lei Tang, Yong Fang,
  and Yang Liu.
\newblock {MalSkillBench}: A runtime-verified benchmark of malicious agent
  skills.
\newblock {\em arXiv preprint arXiv:2606.07131}, 2026.

\bibitem{guo2026skillprobe}
Zihan Guo, Zhiyu Chen, Xiaohang Nie, Jianghao Lin, Yuanjian Zhou, and Weinan
  Zhang.
\newblock {SkillProbe}: Security auditing for emerging agent skill marketplaces
  via multi-agent collaboration.
\newblock {\em arXiv preprint arXiv:2603.21019}, 2026.

\bibitem{hardy1988confused}
Norm Hardy.
\newblock The confused deputy: (or why capabilities might have been invented).
\newblock {\em SIGOPS Oper. Syst. Rev.}, 22(4):36--38, October 1988.
\newblock \href {https://doi.org/10.1145/54289.871709}
  {\path{doi:10.1145/54289.871709}}.

\bibitem{mehedi2025mcp}
Mohammed~Mehedi Hasan, Hao Li, Emad Fallahzadeh, Gopi~Krishnan Rajbahadur, Bram
  Adams, and Ahmed~E. Hassan.
\newblock Model context protocol ({MCP}) at first glance: Studying the security
  and maintainability of {MCP} servers.
\newblock {\em ACM Transactions on Software Engineering and Methodology}, 2026.
\newblock \href {https://doi.org/10.1145/3814959} {\path{doi:10.1145/3814959}}.

\bibitem{he2025pinning}
Hao He, Bogdan Vasilescu, and Christian K{\"a}stner.
\newblock Pinning is futile: You need more than local dependency versioning to
  defend against supply chain attacks.
\newblock {\em Proceedings of the ACM on Software Engineering},
  2(FSE):266--289, 2025.
\newblock \href {https://doi.org/10.1145/3715728} {\path{doi:10.1145/3715728}}.

\bibitem{hou2025mcp}
Xinyi Hou, Yanjie Zhao, Shenao Wang, and Haoyu Wang.
\newblock Model context protocol ({MCP}): Landscape, security threats, and
  future research directions.
\newblock {\em ACM Transactions on Software Engineering and Methodology},
  35(10):1--37, 2026.
\newblock \href {https://doi.org/10.1145/3796519} {\path{doi:10.1145/3796519}}.

\bibitem{hu2024empirical}
Jinchang Hu, Lyuye Zhang, Chengwei Liu, Sen Yang, Song Huang, and Yang Liu.
\newblock Empirical analysis of vulnerabilities life cycle in {Golang}
  ecosystem.
\newblock In {\em Proceedings of the IEEE/ACM 46th International Conference on
  Software Engineering}, pages 1--13, 2024.
\newblock \href {https://doi.org/10.1145/3597503.3639230}
  {\path{doi:10.1145/3597503.3639230}}.

\bibitem{imtiaz2021comparative}
Nasif Imtiaz, Seaver Thorn, and Laurie Williams.
\newblock A comparative study of vulnerability reporting by software
  composition analysis tools.
\newblock In {\em Proceedings of the 15th ACM/IEEE International Symposium on
  Empirical Software Engineering and Measurement (ESEM)}, pages 1--11, 2021.
\newblock \href {https://doi.org/10.1145/3475716.3475769}
  {\path{doi:10.1145/3475716.3475769}}.

\bibitem{jiang2026sok}
Yanna Jiang, Delong Li, Haiyu Deng, Baihe Ma, Xu~Wang, Qin Wang, and Guangsheng
  Yu.
\newblock {SoK}: Agentic skills--beyond tool use in {LLM} agents.
\newblock {\em arXiv preprint arXiv:2602.20867}, 2026.

\bibitem{ladisa2023taxonomy}
Piergiorgio Ladisa, Henrik Plate, Matias Martinez, and Olivier Barais.
\newblock {SoK}: Taxonomy of attacks on open-source software supply chains.
\newblock In {\em 2023 IEEE Symposium on Security and Privacy (SP)}, pages
  1509--1526. IEEE, 2023.
\newblock \href {https://doi.org/10.1109/SP46215.2023.10179304}
  {\path{doi:10.1109/SP46215.2023.10179304}}.

\bibitem{liu2022demystifying}
Chengwei Liu, Sen Chen, Lingling Fan, Bihuan Chen, Yang Liu, and Xin Peng.
\newblock Demystifying the vulnerability propagation and its evolution via
  dependency trees in the npm ecosystem.
\newblock In {\em Proceedings of the 44th International Conference on Software
  Engineering}, pages 672--684, 2022.
\newblock \href {https://doi.org/10.1145/3510003.3510142}
  {\path{doi:10.1145/3510003.3510142}}.

\bibitem{liu2025agentfuzz}
Fengyu Liu, Yuan Zhang, Jiaqi Luo, Jiarun Dai, Tian Chen, Letian Yuan, Zhengmin
  Yu, Youkun Shi, Ke~Li, Chengyuan Zhou, Hao Chen, and Min Yang.
\newblock Make agent defeat agent: Automatic detection of taint-style
  vulnerabilities in {LLM}-based agents.
\newblock In {\em 34th USENIX Security Symposium (USENIX Security 25)}, pages
  3767--3786, 2025.

\bibitem{liu2024llmsmith}
Tong Liu, Zizhuang Deng, Guozhu Meng, Yuekang Li, and Kai Chen.
\newblock Demystifying {RCE} vulnerabilities in {LLM}-integrated apps.
\newblock In {\em Proceedings of the 2024 on ACM SIGSAC Conference on Computer
  and Communications Security}, pages 1716--1730, 2024.
\newblock \href {https://doi.org/10.1145/3658644.3690338}
  {\path{doi:10.1145/3658644.3690338}}.

\bibitem{liu2026maliciousagent}
Yi~Liu, Zhihao Chen, Yanjun Zhang, Gelei Deng, Yuekang Li, Jianting Ning, and
  Leo~Yu Zhang.
\newblock ``{Do} not mention this to the user'': Detecting and understanding
  malicious agent skills in the wild.
\newblock In {\em 35th USENIX Security Symposium (USENIX Security 26)}, pages
  1727--1746, 2026.

\bibitem{liu2026agentwild}
Yi~Liu, Weizhe Wang, Ruitao Feng, Yao Zhang, Guangquan Xu, Gelei Deng, Yuekang
  Li, and Leo Zhang.
\newblock Agent skills in the wild: An empirical study of security
  vulnerabilities at scale.
\newblock {\em arXiv preprint arXiv:2601.10338}, 2026.

\bibitem{livshits2005finding}
V.~Benjamin Livshits and Monica~S. Lam.
\newblock Finding security vulnerabilities in {Java} applications with static
  analysis.
\newblock In {\em 14th USENIX Security Symposium}, 2005.

\bibitem{mitre-cwe}
{MITRE Corporation}.
\newblock {CWE}: Common weakness enumeration.
\newblock \url{https://cwe.mitre.org/}, 2026.
\newblock Accessed: May 2026.

\bibitem{ohm2020backstabber}
Marc Ohm, Henrik Plate, Arnold Sykosch, and Michael Meier.
\newblock Backstabber’s knife collection: A review of open source software
  supply chain attacks.
\newblock In {\em International Conference on Detection of Intrusions and
  Malware, and Vulnerability Assessment}, pages 23--43. Springer, 2020.
\newblock \href {https://doi.org/10.1007/978-3-030-52683-2_2}
  {\path{doi:10.1007/978-3-030-52683-2_2}}.

\bibitem{openai2024gpt4o}
{OpenAI}.
\newblock Hello {GPT-4o}.
\newblock \url{https://openai.com/index/hello-gpt-4o/}, 2024.
\newblock Accessed: May 2026.

\bibitem{openclaw2024}
{OpenClaw}.
\newblock {ClawHub}: Agent skill registry.
\newblock \url{https://clawhub.ai/}, 2024.
\newblock Accessed: May 2026.

\bibitem{pan2026skillguard}
Shidong Pan, Xiaoyu Sun, Tianyi Zhang, Dianshu Liao, Kaiwen Yang, and Zhenchang
  Xing.
\newblock {SkillGuard}: A permission-centric framework for agent skill
  security.
\newblock {\em arXiv preprint arXiv:2606.03024}, 2026.

\bibitem{pashchenko2018vulnerable}
Ivan Pashchenko, Henrik Plate, Serena~Elisa Ponta, Antonino Sabetta, and Fabio
  Massacci.
\newblock Vulnerable open source dependencies: Counting those that matter.
\newblock In {\em Proceedings of the 12th ACM/IEEE International Symposium on
  Empirical Software Engineering and Measurement}, pages 1--10, 2018.
\newblock \href {https://doi.org/10.1145/3239235.3268920}
  {\path{doi:10.1145/3239235.3268920}}.

\bibitem{ponta2018beyond}
Serena~Elisa Ponta, Henrik Plate, and Antonino Sabetta.
\newblock Beyond metadata: Code-centric and usage-based analysis of known
  vulnerabilities in open-source software.
\newblock In {\em 2018 IEEE International Conference on Software Maintenance
  and Evolution (ICSME)}, pages 449--460. IEEE, 2018.
\newblock \href {https://doi.org/10.1109/ICSME.2018.00054}
  {\path{doi:10.1109/ICSME.2018.00054}}.

\bibitem{rahman2025pinning}
Imranur Rahman, Jill Marley, William Enck, and Laurie Williams.
\newblock Which is better for reducing outdated and vulnerable dependencies:
  Pinning or floating?
\newblock In {\em 2025 40th IEEE/ACM International Conference on Automated
  Software Engineering (ASE)}, pages 2795--2807, 2025.
\newblock \href {https://doi.org/10.1109/ASE63991.2025.00229}
  {\path{doi:10.1109/ASE63991.2025.00229}}.

\bibitem{ruan2024toolemu}
Yangjun Ruan, Honghua Dong, Andrew Wang, Silviu Pitis, Yongchao Zhou, Jimmy Ba,
  Yann Dubois, Chris~J. Maddison, and Tatsunori Hashimoto.
\newblock Identifying the risks of {LM} agents with an {LM}-emulated sandbox.
\newblock In {\em The Twelfth International Conference on Learning
  Representations (ICLR)}, 2024.

\bibitem{saha2026skillmd}
Shoumik Saha, Kazem Faghih, and Soheil Feizi.
\newblock Under the hood of {SKILL.md}: Semantic supply-chain attacks on {AI}
  agent skill registry.
\newblock {\em arXiv preprint arXiv:2605.11418}, 2026.

\bibitem{schmotz2026skill}
David Schmotz, Luca Beurer-Kellner, Sahar Abdelnabi, and Maksym Andriushchenko.
\newblock Skill-inject: Measuring agent vulnerability to skill file attacks.
\newblock {\em arXiv preprint arXiv:2602.20156}, 2026.

\bibitem{vu2020typosquatting}
Duc-Ly Vu, Ivan Pashchenko, Fabio Massacci, Henrik Plate, and Antonino Sabetta.
\newblock Typosquatting and combosquatting attacks on the {Python} ecosystem.
\newblock In {\em 2020 IEEE European Symposium on Security and Privacy
  Workshops (EuroS\&PW)}, pages 509--514. IEEE, 2020.
\newblock \href {https://doi.org/10.1109/EuroSPW51379.2020.00074}
  {\path{doi:10.1109/EuroSPW51379.2020.00074}}.

\bibitem{xi2025agents}
Zhiheng Xi, Wenxiang Chen, Xin Guo, Wei He, Yiwen Ding, Boyang Hong, Ming
  Zhang, Junzhe Wang, Senjie Jin, Enyu Zhou, et~al.
\newblock The rise and potential of large language model based agents: A
  survey.
\newblock {\em Science China Information Sciences}, 68(2):121101, 2025.
\newblock \href {https://doi.org/10.1007/s11432-024-4222-0}
  {\path{doi:10.1007/s11432-024-4222-0}}.

\bibitem{yamaguchi2014modeling}
Fabian Yamaguchi, Nico Golde, Daniel Arp, and Konrad Rieck.
\newblock Modeling and discovering vulnerabilities with code property graphs.
\newblock In {\em 2014 IEEE Symposium on Security and Privacy}, pages 590--604,
  2014.

\bibitem{zahan2022weak}
Nusrat Zahan, Thomas Zimmermann, Patrice Godefroid, Brendan Murphy, Chandra
  Maddila, and Laurie Williams.
\newblock What are weak links in the npm supply chain?
\newblock In {\em Proceedings of the 44th International Conference on Software
  Engineering: Software Engineering in Practice}, pages 331--340, 2022.
\newblock \href {https://doi.org/10.1145/3510457.3513044}
  {\path{doi:10.1145/3510457.3513044}}.

\bibitem{zhan2025adaptive}
Qiusi Zhan, Richard Fang, Henil~Shalin Panchal, and Daniel Kang.
\newblock Adaptive attacks break defenses against indirect prompt injection
  attacks on {LLM} agents.
\newblock In {\em Findings of the Association for Computational Linguistics:
  NAACL 2025}, pages 7101--7117, 2025.
\newblock \href {https://doi.org/10.18653/v1/2025.findings-naacl.395}
  {\path{doi:10.18653/v1/2025.findings-naacl.395}}.

\bibitem{zhan2024injecagent}
Qiusi Zhan, Zhixiang Liang, Zifan Ying, and Daniel Kang.
\newblock Injecagent: Benchmarking indirect prompt injections in
  tool-integrated large language model agents.
\newblock In {\em Findings of the Association for Computational Linguistics:
  ACL 2024}, pages 10471--10506, 2024.
\newblock \href {https://doi.org/10.18653/v1/2024.findings-acl.624}
  {\path{doi:10.18653/v1/2024.findings-acl.624}}.

\bibitem{zhang2023mitigating}
Lyuye Zhang, Chengwei Liu, Sen Chen, Zhengzi Xu, Lingling Fan, Lida Zhao, Yiran
  Zhang, and Yang Liu.
\newblock Mitigating persistence of open-source vulnerabilities in maven
  ecosystem.
\newblock In {\em 2023 38th IEEE/ACM International Conference on Automated
  Software Engineering (ASE)}, pages 191--203. IEEE, 2023.
\newblock \href {https://doi.org/10.1109/ASE56229.2023.00058}
  {\path{doi:10.1109/ASE56229.2023.00058}}.

\bibitem{zhu2026skillclone}
Jiaying Zhu, Lyuye Zhang, Wenbo Guo, and Yang Liu.
\newblock Latent reuse in agent skills: Multi-modal clone detection at
  ecosystem scale.
\newblock In {\em Proceedings of the 41st IEEE/ACM International Conference on
  Automated Software Engineering}, 2026.
\newblock \href {https://doi.org/10.1145/3832783.3837468}
  {\path{doi:10.1145/3832783.3837468}}.

\bibitem{zimmermann2019npm}
Markus Zimmermann, Cristian-Alexandru Staicu, Cam Tenny, and Michael Pradel.
\newblock Small world with high risks: A study of security threats in the npm
  ecosystem.
\newblock In {\em 28th USENIX Security symposium (USENIX security 19)}, pages
  995--1010, 2019.

\end{thebibliography}
